\documentclass[english,prb,twocolumn,citeautoscript]{revtex4-1}
\usepackage[T1]{fontenc}
\usepackage[latin9]{inputenc}
\usepackage{color}
\usepackage{float}
\usepackage{units}
\usepackage{amstext}
\usepackage{graphicx}

\makeatletter

\providecommand{\tabularnewline}{\\}

\setcitestyle{super}

\@ifundefined{showcaptionsetup}{}{%
 \PassOptionsToPackage{caption=false}{subfig}}
\usepackage{subfig}
\makeatother

\usepackage{babel}
\begin{document}

\title{{\normalsize{}Optical signatures of electric field driven magnetic
phase transitions in graphene quantum dots}}

\author{{\normalsize{}{*}Tista Basak and $^{\$}$Alok Shukla}}

\affiliation{{*}Mukesh Patel School of Technology Management and Engineering,
NMIMS University, Mumbai-56, India}

\affiliation{$^{\$}$Department of Physics, Indian Institute of Technology Bombay,
Powai, Mumbai-400076, INDIA.}
\email{tista.basak@nmims.edu, shukla@phy.iitb.ac.in}

\begin{abstract}
{\normalsize{}Experimental challenges in identifying various types
of magnetic ordering in graphene quantum dots (QDs) pose a major hurdle
in the application of these nanostructures for spintronic devices.
}Based upon phase diagrams obtained by employing {\normalsize{}the
$\pi$-electron }Pariser-Parr-Pople (PPP) model Hamiltonian, we demonstrate
that the magnetic states undergo phase transition under the influence
of an external electric field. Our calculations of the electro-absorption
spectra of these QDs indicate that the spectrum in question carries
strong signatures of their magnetic state (FM vs AFM), thus suggesting
the possibility of an all-optical characterization of their magnetic
nature. Further, the gaps for the up and the down spins are the same
in the absence of an external electric field, both for the antiferromagnetic
(AFM), and the ferromagnetic (FM) states of QDs. But, once the QDs
are exposed to a suitably directed external electric field, gaps for
different spins split, and, exhibit distinct variations with respect
to the strength of the field. The nature of variation exhibited by
the energy gaps corresponding to the up and down spins is different
for the AFM and FM configurations of QDs. This selective manipulation
of the spin-polarized gap splitting by an electric field in finite
graphene nanostructures can open up new frontiers in the design of
graphene-based spintronic devices. 
\end{abstract}
\maketitle

\section{Introduction}

The existence of intrinsic magnetism in graphene is a highly controversial
issue which has evoked considerable research interest among the scientific
fraternity for the past few years\citep{Son,PhysRevB.82.201411Agapito,ZhengdopedPhysRevB.78.155118,Bhowmick:/content/aip/journal/jcp/128/24/10.1063/1.2943678,:/content/aip/journal/apl/103/13/10.1063/1.4821954Zhou,PhysRevB.86.045449Ma}.
Theoretical studies\citep{Son,:/content/aip/journal/apl/103/13/10.1063/1.4821954Zhou,:/content/aip/journal/jap/108/7/10.1063/1.3489919Sahin,Bhowmick:/content/aip/journal/jcp/128/24/10.1063/1.2943678,doi:10.1021/nl072548aWang,FeldnerPhysRevB.81.115416,PhysRevB.77.235411Zhang,PhysRevB.78.045421Zheng,PhysRevB.82.201411Agapito,PhysRevB.84.245403Zarenia,PhysRevB.86.045449Ma,PhysRevB.87.035425Guclu,PhysRevLett.99.177204Fernandes,Agapitodoi:10.1021/jp1096234,efieldmag-hawrylak,efieldmag-karol}
have revealed that quantum confinement, shape, edge topology, and
application of an external electric field have a profound influence
on the magnetic properties of graphene-based nanostructures which
can be efficiently exploited for designing novel spintronic devices\citep{Tombros,PhysRevLett.100.047209Yazyev,:/content/aip/journal/apl/94/22/10.1063/1.3147203Han,PhysRevLett.107.047206Yang}.
It has been recently predicted that the application of electric field
can induce energy-level shifts of spin-ordered edge states resulting
in phase transition between the different magnetic states of zigzag-edged
graphene nanostructures.\citep{Son,ZhengdopedPhysRevB.78.155118,PhysRevB.82.201411Agapito,PhysRevB.86.045449Ma,:/content/aip/journal/apl/103/13/10.1063/1.4821954Zhou}
Further, it was revealed that in case of the antiferromagnetic (AFM)
phase of graphene nanostructures, the applied electric field breaks
the spin degeneracy leading to half-semiconducting behavior.\citep{ZhengdopedPhysRevB.78.155118}
Although a number of theoretical calculations probing magnetism in
regular shaped graphene quantum dots (QDs) have been performed, very
little literature exists on such studies of irregular shaped QDs.\citep{Chen_nature}
Recently, Chen \emph{et al.},\citep{Chen_nature} provided experimental
evidence of intrinsic magnetism in graphene sheets with irregular
zigzag edges. However, indisputable experimental evidence to corroborate
various theoretical predictions is still lacking due to the complications
involved in precision measurements of weak magnetic signals in graphene
nanostructures by employing current techniques \citep{Chen_nature,Magda,PhysRevB.84.045421_Kiguchi,Suenaga_nature}.
This has inspired us to look at other options to solve this problem,
and in an earlier work\citep{PhysRevB.83.075413_Gundra}, we had analyzed
an all-optical technique based upon electroabsorption (EA) spectra
to efficiently detect the magnetic ground state of one-dimensional
structures, \emph{viz}., graphene nanoribbons. In this work, we investigate
the use of EA spectroscopy to probe different magnetic configurations
of zero-dimensional graphene structures, i.e., QDs. Employing a correlated
$\pi$-electron approach, we compute the linear optical response of
graphene QDs of various shapes and sizes with, and without, static
external electric fields applied in the plane of QDs in various directions,
and find that it is highly dependent upon the magnetic states of the
QDs concerned. In particular, the nature of variation exhibited by
the energy gaps corresponding to the up and down spins is different
for the AFM and FM configurations of QDs. Therefore, we argue that
the EA spectroscopy can prove to be a powerful alternative to convention
magnetic measurements for determining the magnetic states of graphene
QDs. Furthermore, selective manipulation of the spin-dependent splitting
of gaps by an electric field in finite graphene nanostructures, namely,
graphene QDs, can open up new frontiers in the design of graphene-based
spintronic devices.

Remainder of this paper is organized as follows. In section II we
briefly describe our theoretical methodology, while in section III
our results are presented and discussed. Finally, in section IV we
present our conclusions.

\section{Theoretical \label{sec:Theoretical-Methodology}Methodology}

\begin{figure}
\subfloat[BQD-38]{\includegraphics[width=1.4cm]{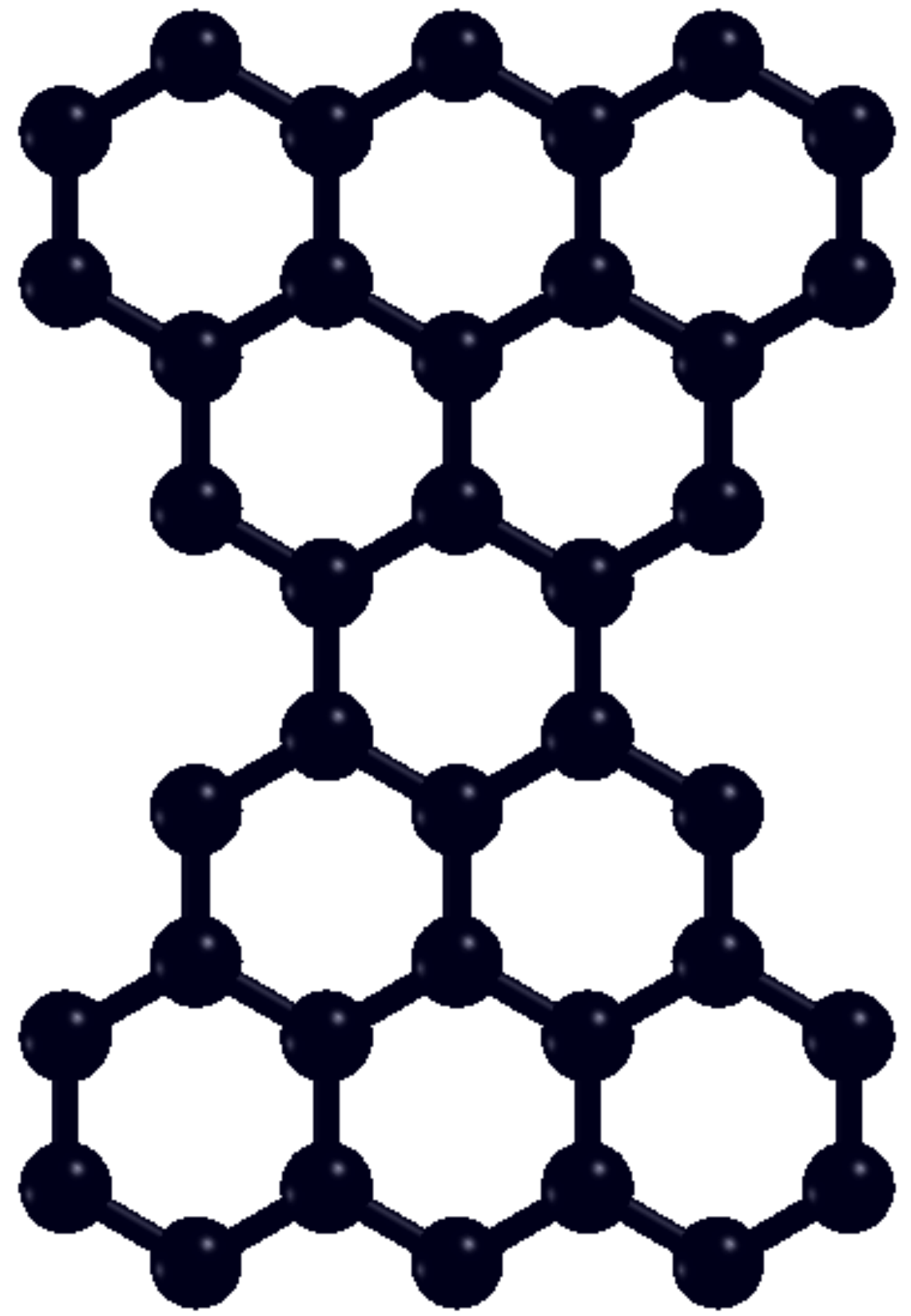}

}\subfloat[RQD-54]{

\includegraphics[width=1.4cm]{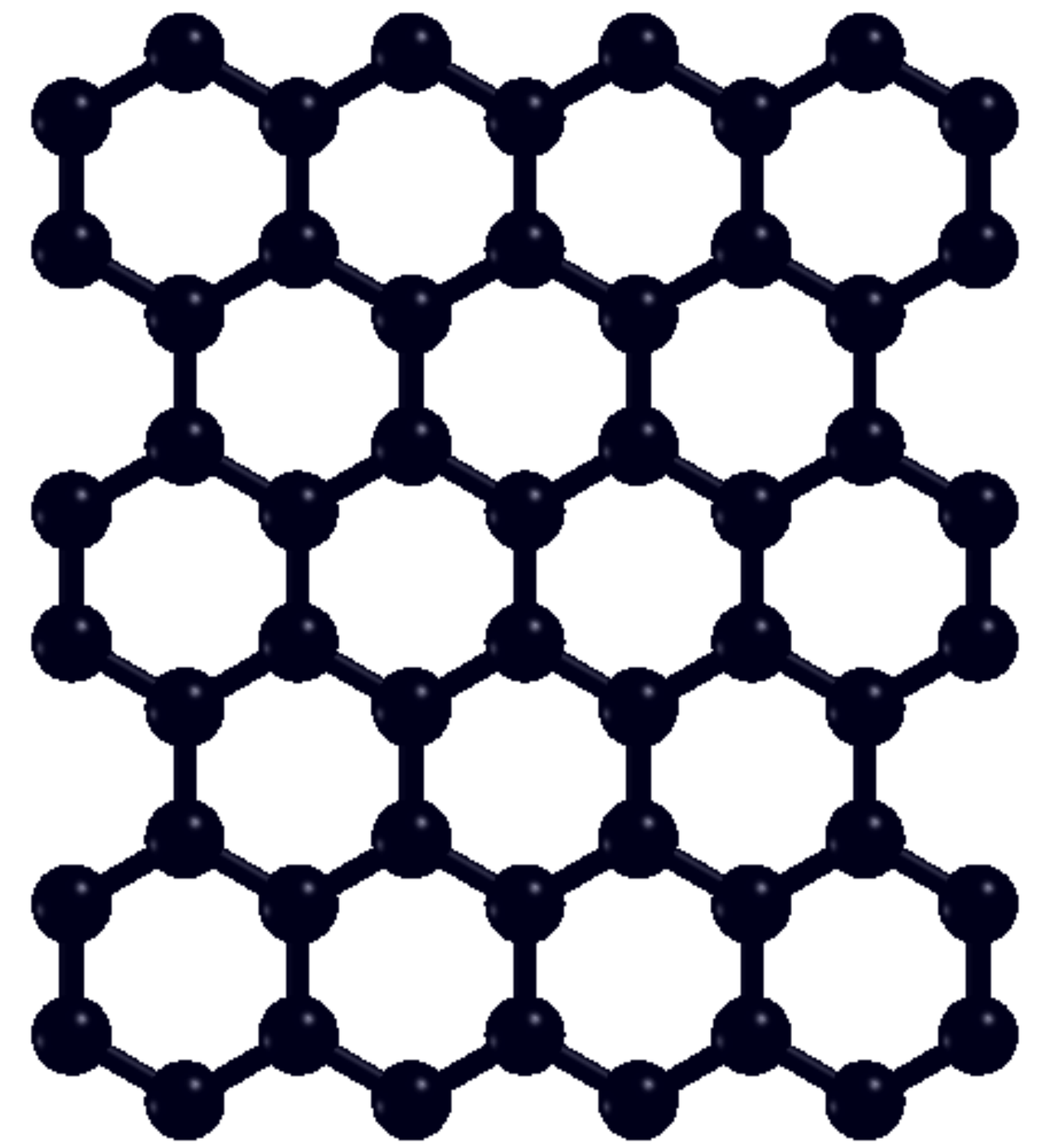}

}\subfloat[GQD-38]{

\includegraphics[width=1.5cm]{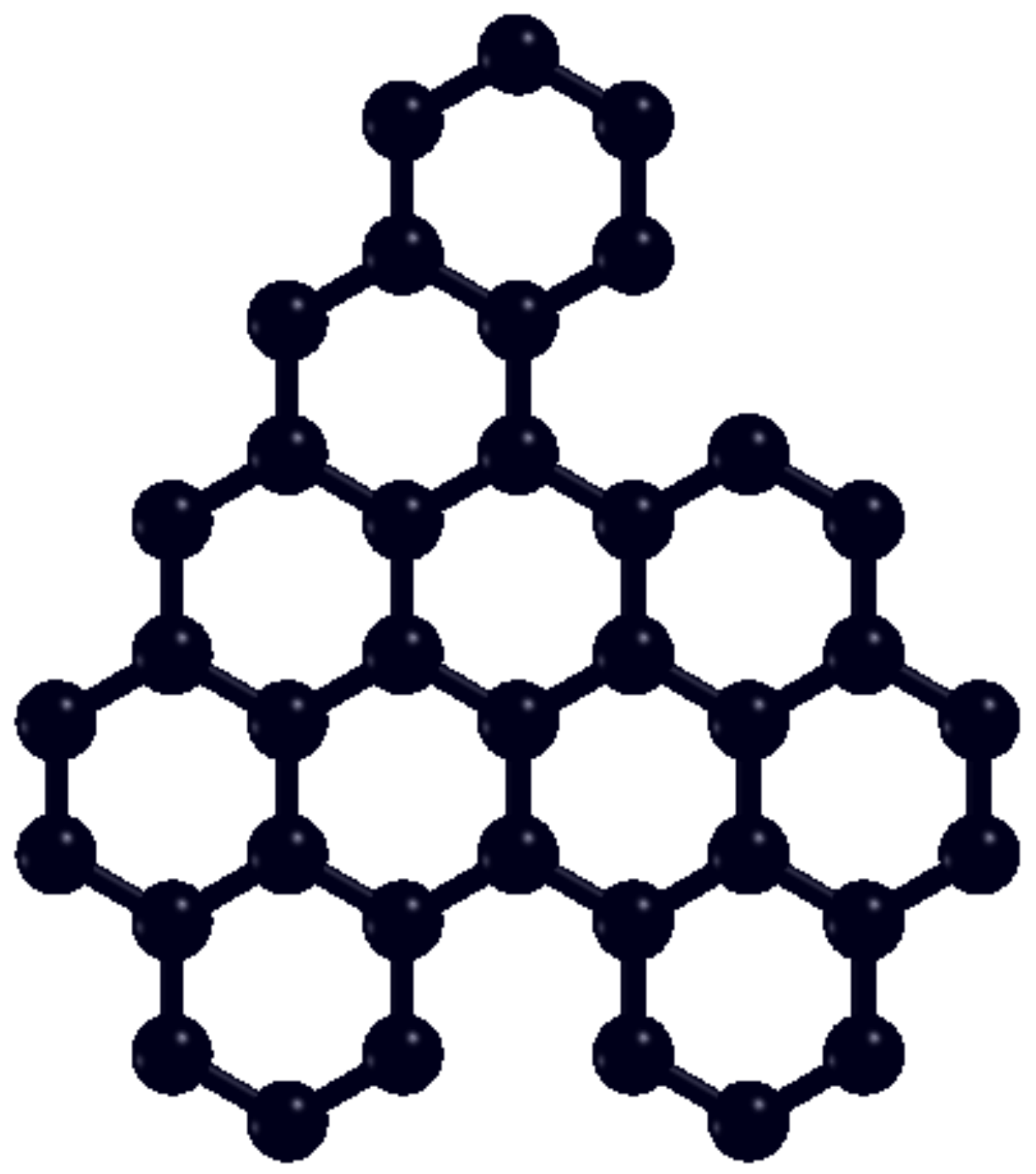}

}\subfloat[GQD-40]{

\includegraphics[width=1.5cm]{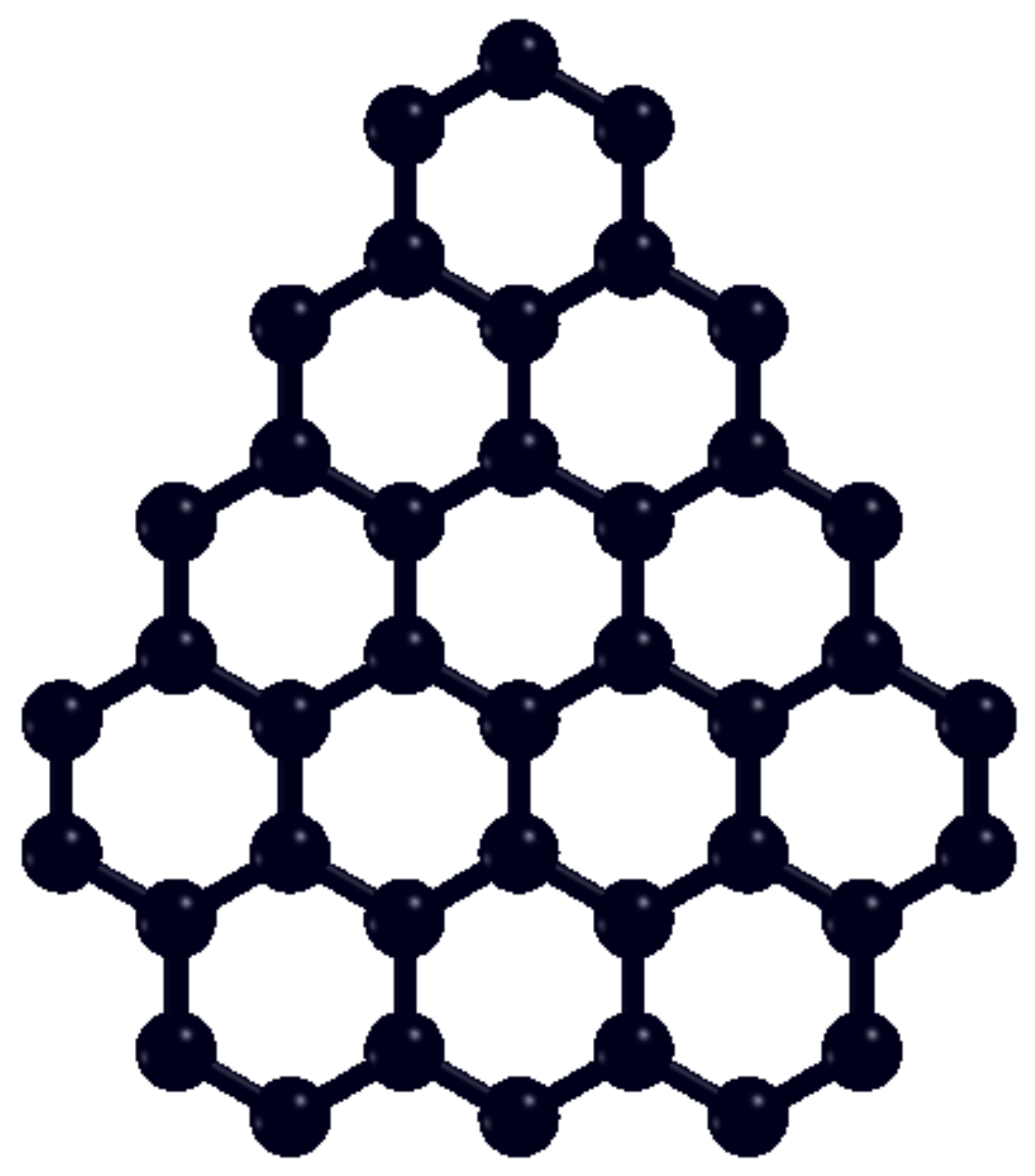}

}\subfloat[GQD-48]{

\includegraphics[width=1.5cm]{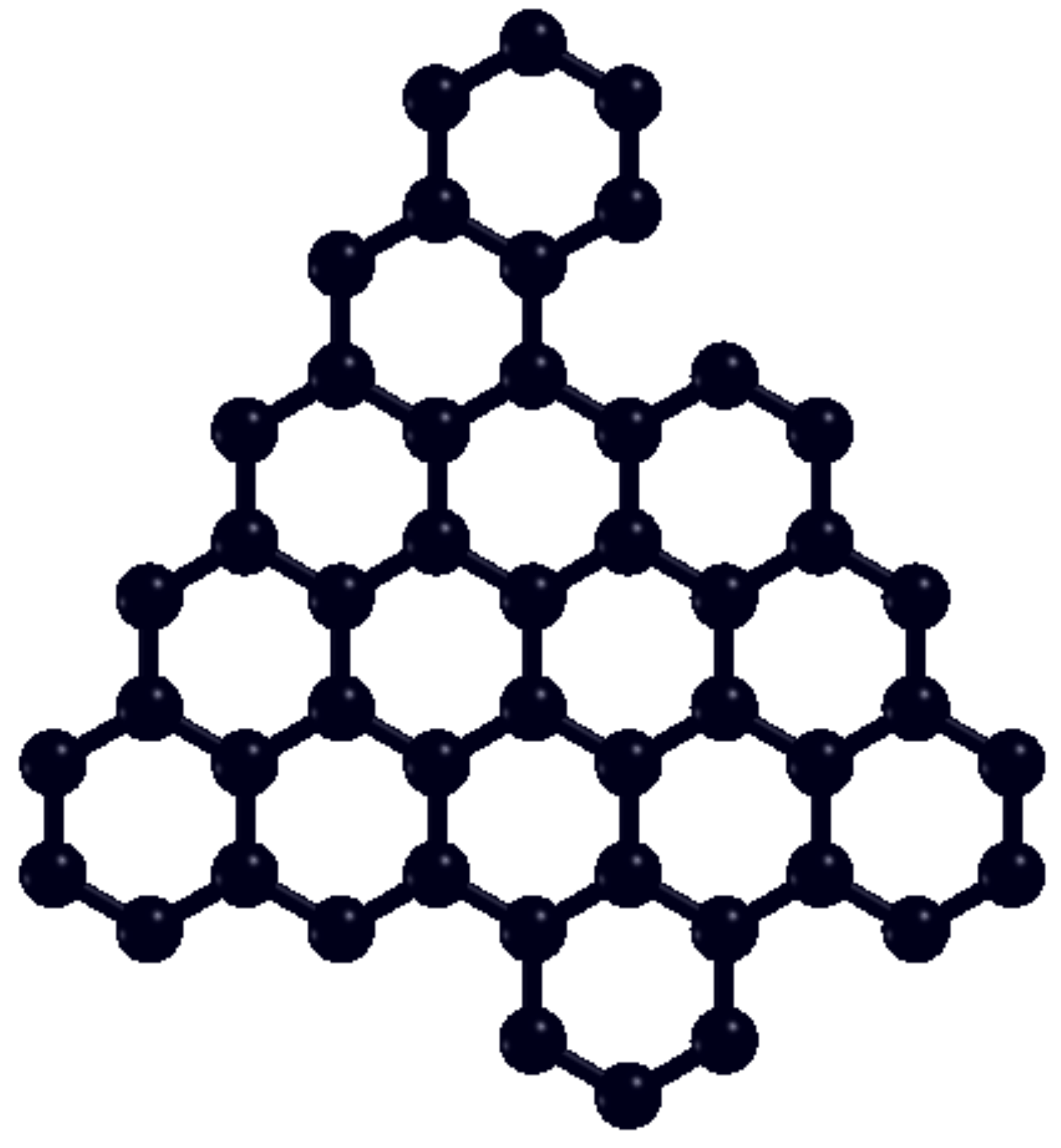}

}

\caption{Schematic \label{fig:Schematic-diagram-of}diagrams of QDs considered
in this work: (a) bowtie shaped (BQD-38), (b) rectangle shaped (RQD-54),
while (c), (d), and (e) denote dots of lower symmetry GQD-38, GQD-40,
and GQD-48, respectively.}
\end{figure}

The symmetric structures considered here (\emph{cf.} Fig. \ref{fig:Schematic-diagram-of})
include a rectangular QD with 54 carbon atoms (RQD-54), and a bowtie
shaped QD with 38 atoms (BQD-38), both of which have $D_{2h}$ symmetry.
The QD with 40 atoms (GQD-40) exhibits lower $C_{2v}$ symmetry, while
those with 38, and 48 atoms (GQD-38 and GQD-48) are completely asymmetric.
Quantum dots RQD-54, BQD-38, and GQD-38 have balanced sub-lattices,
while GQD-40 and GQD-48 have imbalanced sub-lattices. 

These calculations have been carried out using the effective $\pi$-electron
Pariser-Parr-Pople (PPP) model Hamiltonian,\citep{ppp-pople,ppp-pariser-parr} 

\begin{eqnarray}
H & \ensuremath{=} & \mbox{\mbox{\mbox{\mbox{\ensuremath{-}\ensuremath{\sum_{i,j,\sigma}t_{ij}\left(c_{i\sigma}^{\dagger}c_{j\sigma}+c_{j\sigma}^{\dagger}c_{i\sigma}\right)}\ensuremath{+}\ensuremath{U\sum_{i}n_{i\uparrow}n_{i\downarrow}}}}}}\nonumber \\
 &  & \ensuremath{+}\ensuremath{\sum_{i<j}V_{ij}(n_{i}-1)(n_{j}-1)}\label{eq:ppp}
\end{eqnarray}

where $c_{i\sigma}^{\dagger}($c$_{i\sigma})$ creates (annihilates)
a $\pi$ orbital of spin $\sigma$, localized on the \emph{i}th carbon
atom, while the total number of electrons with spin $\sigma$ on atom
$i$ is indicated by $n$$_{i}=\sum_{\sigma}c_{i\sigma}^{\dagger}c_{i\sigma}$.
Further, $t_{ij}$, $U$, and $V_{ij}$, denote hopping, onsite Coulomb
repulsion, and long-range Coulomb interactions, respectively. \textcolor{black}{Hopping
matrix elements $t_{ij}$ were restricted to nearest neighbor sites
$i$ and $j$, with their uniform value $t$$_{0}$ taken to be 2.4
eV, consistent with our earlier calculations on polymers}\citep{PhysRevB.71.165204Priya_t0},
\textcolor{black}{polycyclic aromatic hydrocarbons}\citep{:/content/aip/journal/jcp/140/10/10.1063/1.4867363Aryanpour},
\textcolor{black}{and hydrogenated graphene nanofragments}\citep{Tista}.
\textcolor{black}{Coulomb interaction in the PPP model Hamiltonian
are parametrized as per the Ohno relationship}\citep{Theor.chim.act.2Ohno}

\begin{equation}
V_{ij}=U/\kappa_{i,j}(1+0.6117R_{i,j}^{2})^{\nicefrac{1}{2}},
\end{equation}

\textcolor{black}{where $U$, as described above, is the on-site electron-electron
repulsion term, $\kappa_{i,j}$ represents the dielectric constant
of the system which simulates the screening effects, and $R$$_{i,j}$
is the distance} (in \AA) \textcolor{black}{between the $i$th and
$j$th carbon atoms. In this work, we have performed calculations
adopting ``screened parameters'', with $U$ = 8.0 eV, $\kappa_{i,j}$=
2.0 $(i$$\neq j)$, proposed initially by Chandross and Mazumdar,
for studying the optical absorption in PPV}\citep{PhysRevB.55.1497Chandross},
\textcolor{black}{and also used in several of our earlier works}\citep{:/content/aip/journal/jcp/131/1/10.1063/1.3159670Priyaanthracene,:/content/aip/journal/jcp/140/10/10.1063/1.4867363Aryanpour,doi10.1021/jp408535u,doi:10.1021/jp410793rAryanpour,himanshu-triplet,PhysRevB.65.125204Shukla65,PhysRevB.69.165218Shukla69,PhysRevB.71.165204Priya_t0,PhysRevB.83.075413_Gundra,PhysRevB.84.075442Gundra,sony-acene-lo,Sony2010821}.
\textcolor{black}{Present calculations were performed at the restricted
Hartree-Fock (RHF) level for the non-magnetic states, and the unrestricted
Hartree-Fock (UHF) level for the magnetic states, using a code developed
in our group} \citep{Sony2010821}. \textcolor{black}{The details
of the calculations are given extensively in our earlier works}\citep{Sony2010821,Tista,:/content/aip/journal/jcp/140/10/10.1063/1.4867363Aryanpour,PhysRevB.83.075413_Gundra,PhysRevB.84.075442Gundra}. 

All QDs considered here are assumed to lie in the $x$-$y$ plane,
with their longer dimension along the $y$ axis. For EA calculations,
the electric field is applied along the $y$-axis (transverse direction),
$x$-axis (longitudinal direction), and diagonal directions in the
$x$-$y$ plane. All carbon-carbon bond lengths and bond angles have
been fixed at 1.4 \AA, and 120\textdegree, respectively. 

\section{Results and Discussion\label{sec:Results-and-Discussion}}

\subsection{Ordering of states and HOMO-LUMO band gaps in the absence of electric
field}

In Table \ref{tab:HOMO-LUMO-band-gap-and-1} we present the calculated
difference in Hartree Fock (HF) total energy of the excited states,
with respect to the ground state, and the HOMO-LUMO (H-L) band gaps
for different magnetic phases of graphene QDs, in the absence of electric
field. For the FM (AFM) phases, the $z$-component of the total spin
($S_{z}$) was taken to be 1 (0). 

\begin{table}
\begin{tabular}{|c|ccc|ccc|}
\hline 
QD & HF energy & difference &  & H-L & band-gap  & \tabularnewline
\hline 
 & E$_{exc}$- E$_{gnd}$ & (eV) &  &  & (eV) & \tabularnewline
\hline 
 & AFM & FM & NM & AFM  & FM  & NM \tabularnewline
 & phase & phase & phase & phase & phase & phase\tabularnewline
\hline 
RQD-54 & 0.00 & 0.47 & 1.79 & 3.97 & 2.76 & 0.90\tabularnewline
\hline 
BQD-38 & 0.00 & 0.49 & 4.72 & 4.24 & 3.25 & 0.58\tabularnewline
\hline 
GQD-38 & 0.00 & 0.88 & 0.59 & 4.02 & 2.94 & 2.47\tabularnewline
\hline 
GQD-40 & 0.83 & 0.00 & 2.01 & 2.47 & 4.39 & 0.55\tabularnewline
\hline 
GQD-48 & 0.78 & 0.00 & 2.08 & 2.49 & 4.03 & 0.51\tabularnewline
\hline 
\end{tabular}\caption{Calculated difference in Hartree-Fock energy of the excited states
with respect to the ground state and HOMO-LUMO \label{tab:HOMO-LUMO-band-gap-and-1}band-gap
for different magnetic phases of graphene QDs, in the absence of electric
field. For the FM (AFM) phases $z$-component of the total spin ($S_{z}$)
was taken to be 1 (0). For a given QD, $\mbox{E}_{\mbox{exc}}$($\mbox{E}_{\mbox{gnd}}$),
denotes the HF total energy of its higher (lower) energy magnetic
state (AFM/FM)}
\end{table}

An analysis of the energy difference (Table \ref{tab:HOMO-LUMO-band-gap-and-1})
indicates that the QDs with balanced sublattices, i.e., equal number
of $A$ and $B$ type atoms (RQD-54, BQD-38 and GQD-38), have the
AFM state as the ground state, while the ones with imbalanced sublattices,
namely, GQD-40 and GQD-48, have a ferromagnetic (FM) ground state.
This is consistent with the\textcolor{red}{{} }Lieb's theorem\citep{Liebs-theorem}
which states that the spin S of the ground state of the Hubbard model
in neutral bipartite lattices is given by $2S$$=N_{A}-N_{B}$ , where
$N_{A}$ and $N_{B}$ represent the number of atoms constituting each
sublattice. As far as the energetic ordering of various magnetic states
is concerned, for RQD-54 and BQD-38, FM state follows the AFM ground
state, with the non-magnetic (NM) state placed the highest. In comparison,
for GQD-38, the excited state ordering is reversed. In case of GQD-40
and GQD-48, the AFM state appears next after the ground state, followed
by the NM configuration. 

The calculated H-L band gap (Table \ref{tab:HOMO-LUMO-band-gap-and-1})
is largest for the ground-state configuration (namely AFM phase for
RQD-54, BQD-38 and GQD-38 and FM state for GQD-40 and GQD-48) while
it is lowest for the NM state for all the QDs. Thus, it is possible
to identify the ground-state magnetic coupling by analyzing the H-L
band gap (optical band gap) of these QDs.

\subsection{Spin-density plots of different magnetic phases of graphene quantum
dots in the absence of electric field}

\subsubsection{AFM phase}

\begin{figure}
\subfloat[BQD-38]{\includegraphics[width=1.4cm]{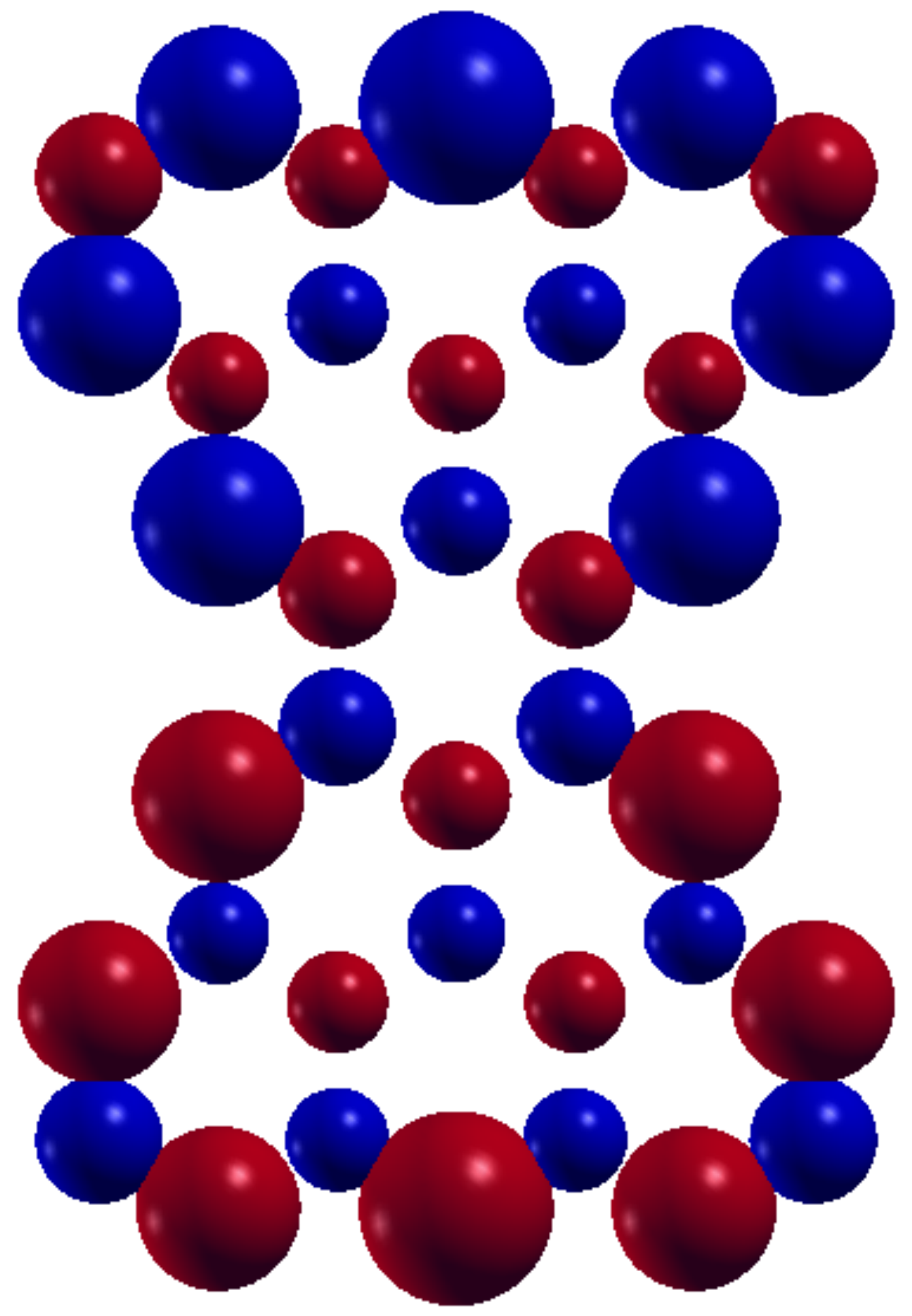}

}\subfloat[RQD-54]{\includegraphics[width=1.4cm]{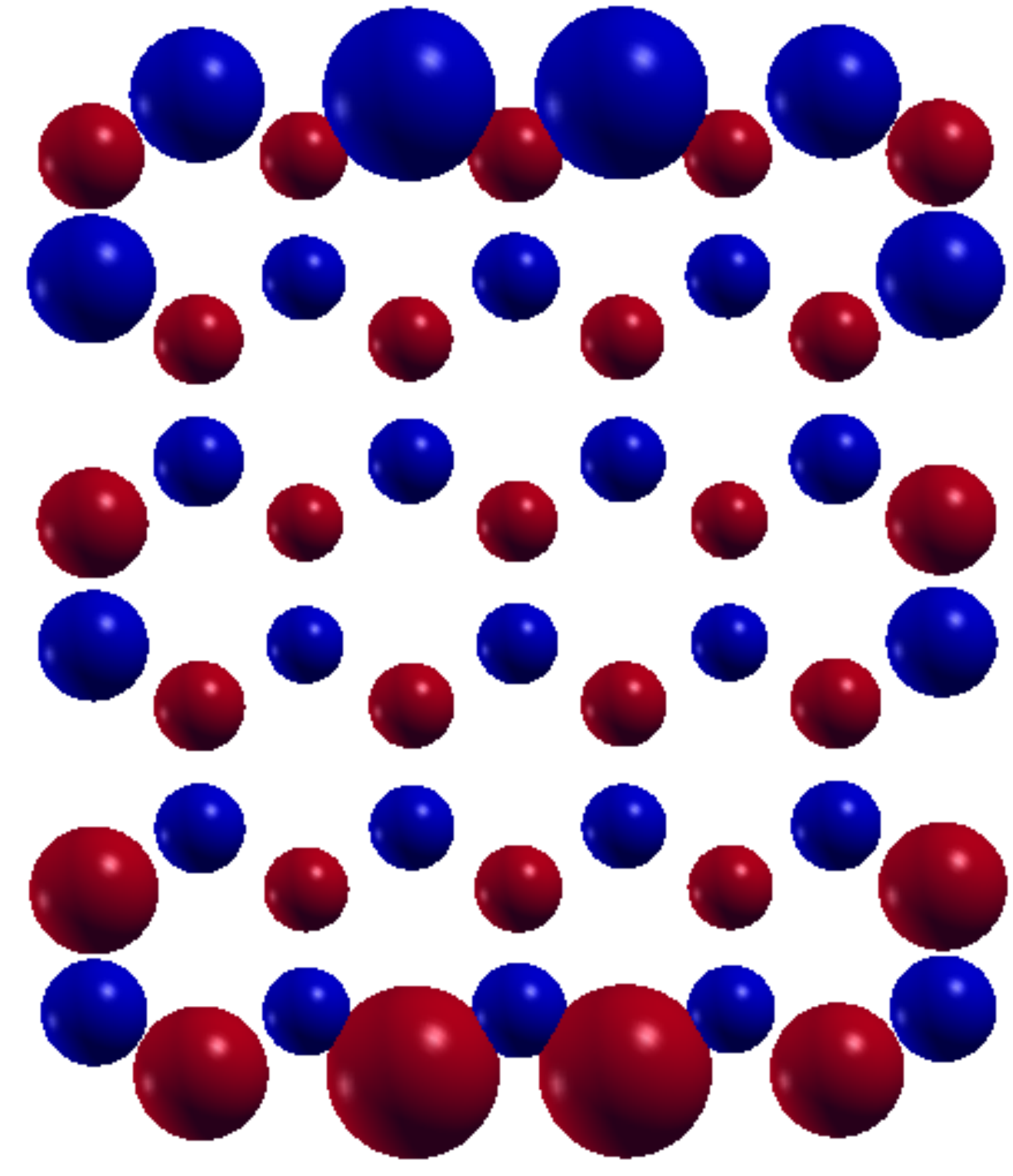}}\subfloat[GQD-38]{\includegraphics[width=1.4cm]{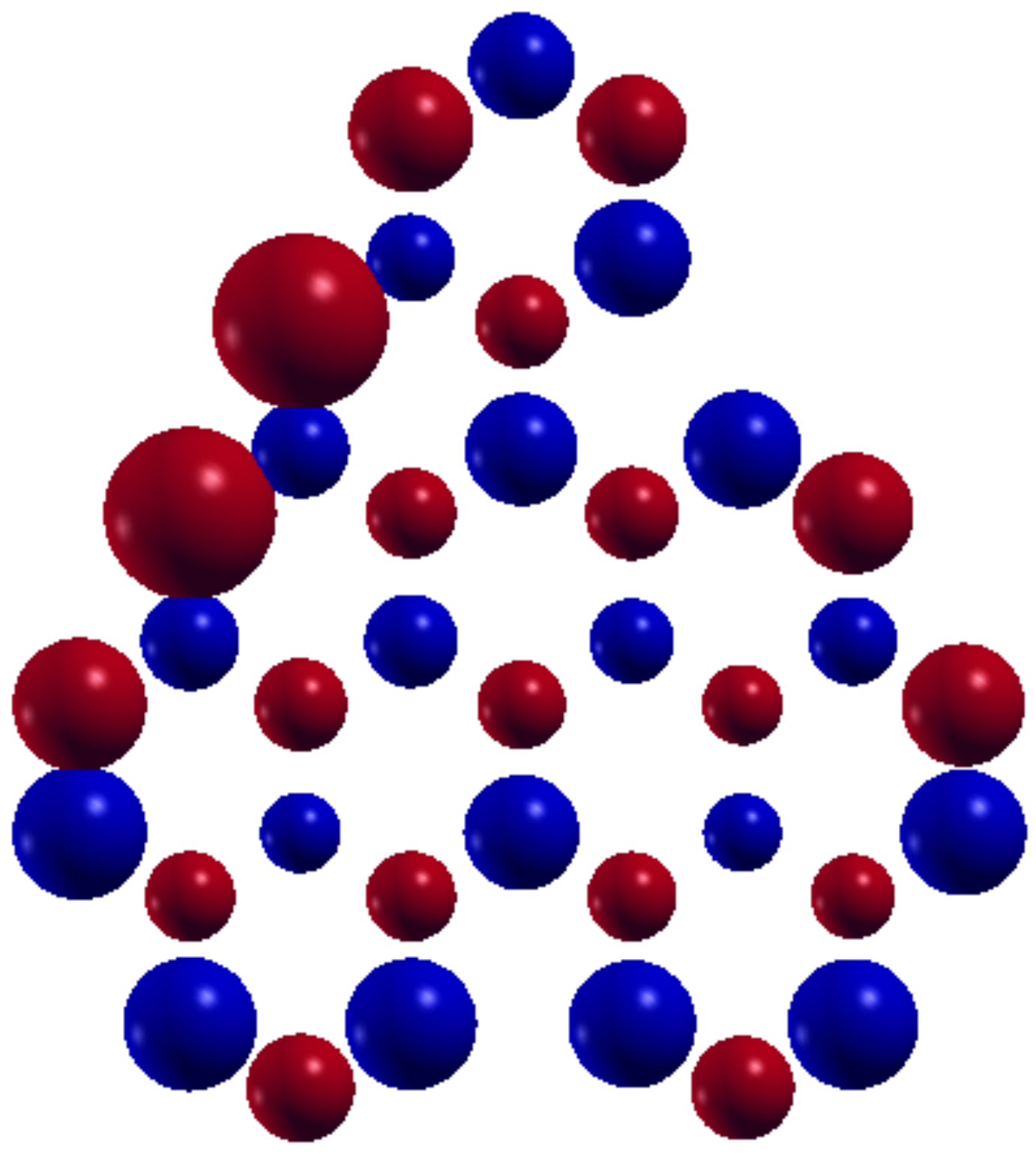}

}\subfloat[GQD-40]{\includegraphics[width=1.4cm]{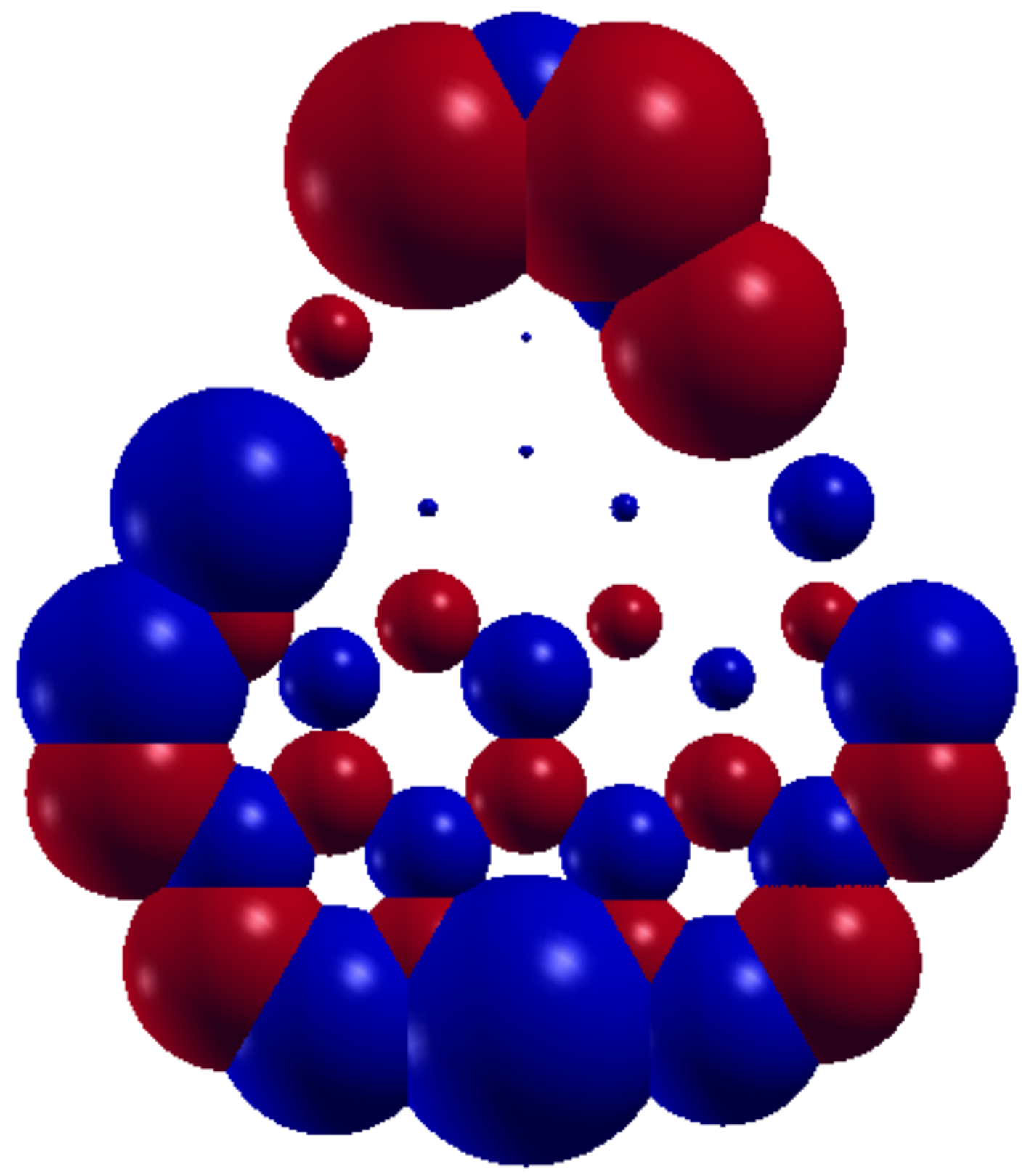}}\subfloat[GQD-48]{\includegraphics[width=1.4cm]{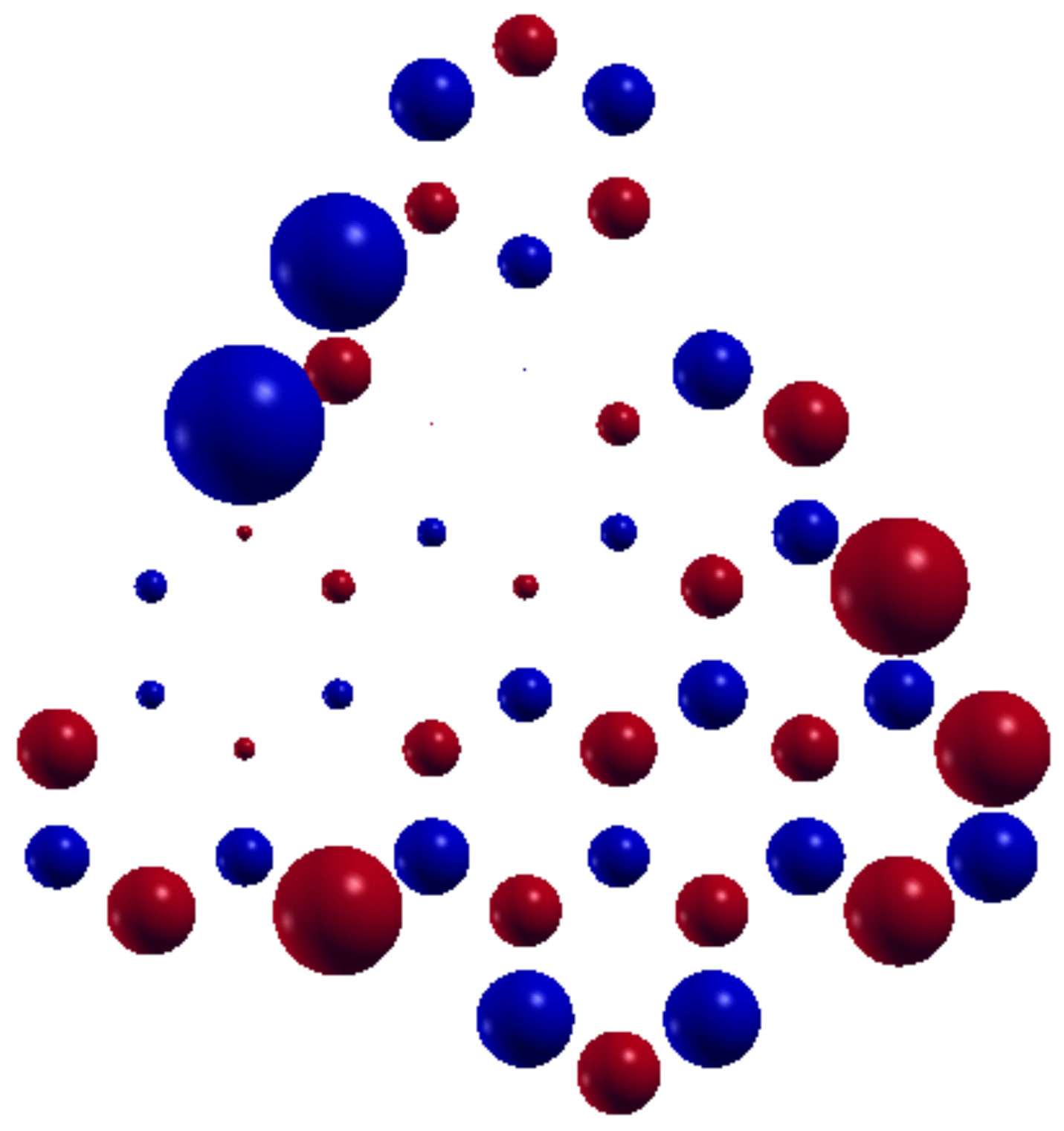}}\caption{Spin-density plots of AFM configurations of: (a) BQD-38, (b) RQD-54,
(c) GQD-38, (d) GQD-40, and (e) GQD-48, in the absence of electric
field. The red and the blue spheres represent the two different spin
orientations (up/down or $\alpha/\beta$) of the carbon atoms. \label{fig:Spin-density-plots-of-AFM}}
\end{figure}

Figure \ref{fig:Spin-density-plots-of-AFM} represents spin-density
plots of AFM configuration of BQD-38, RQD-54, GQD-38, GQD-40 and GQD-48,
in the absence of electric field. The red and blue spheres represent
the two different spin orientations (up/down or $\alpha/\beta$) of
the carbon atoms. 

It is obvious from the figure that the spin densities corresponding
to the two different spin orientations are localized on the opposite
sides of the different quantum dots under consideration. This spatial
asymmetry of the spin densities gives rise to local magnetism with
zero net spin, expected for the AFM case.

\subsubsection{FM phase}

\begin{figure}[h]
\subfloat[BQD-38]{\includegraphics[width=1.5cm]{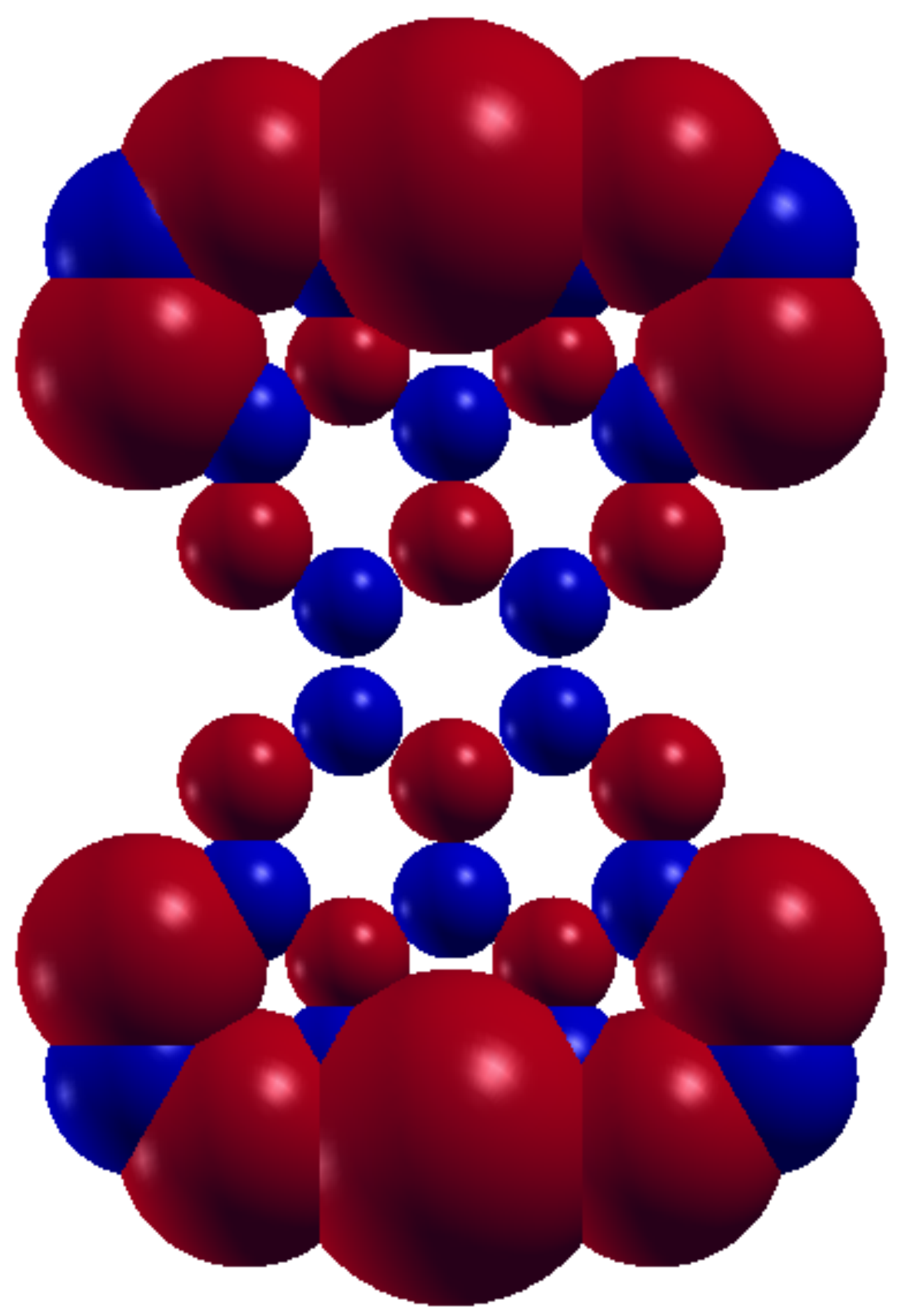}}\subfloat[RQD-54]{\includegraphics[width=1.4cm]{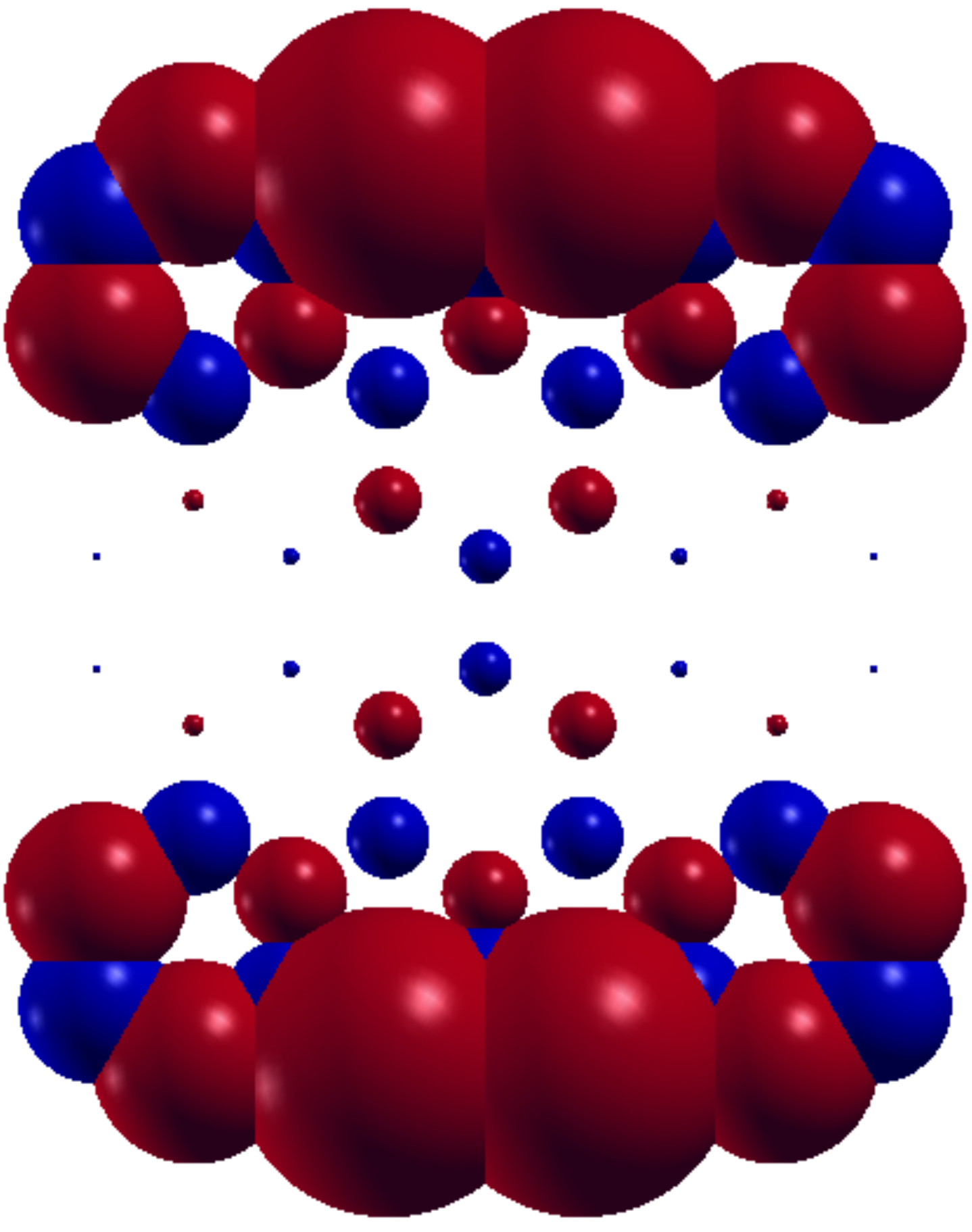}}\subfloat[GQD-38]{\includegraphics[width=1.4cm]{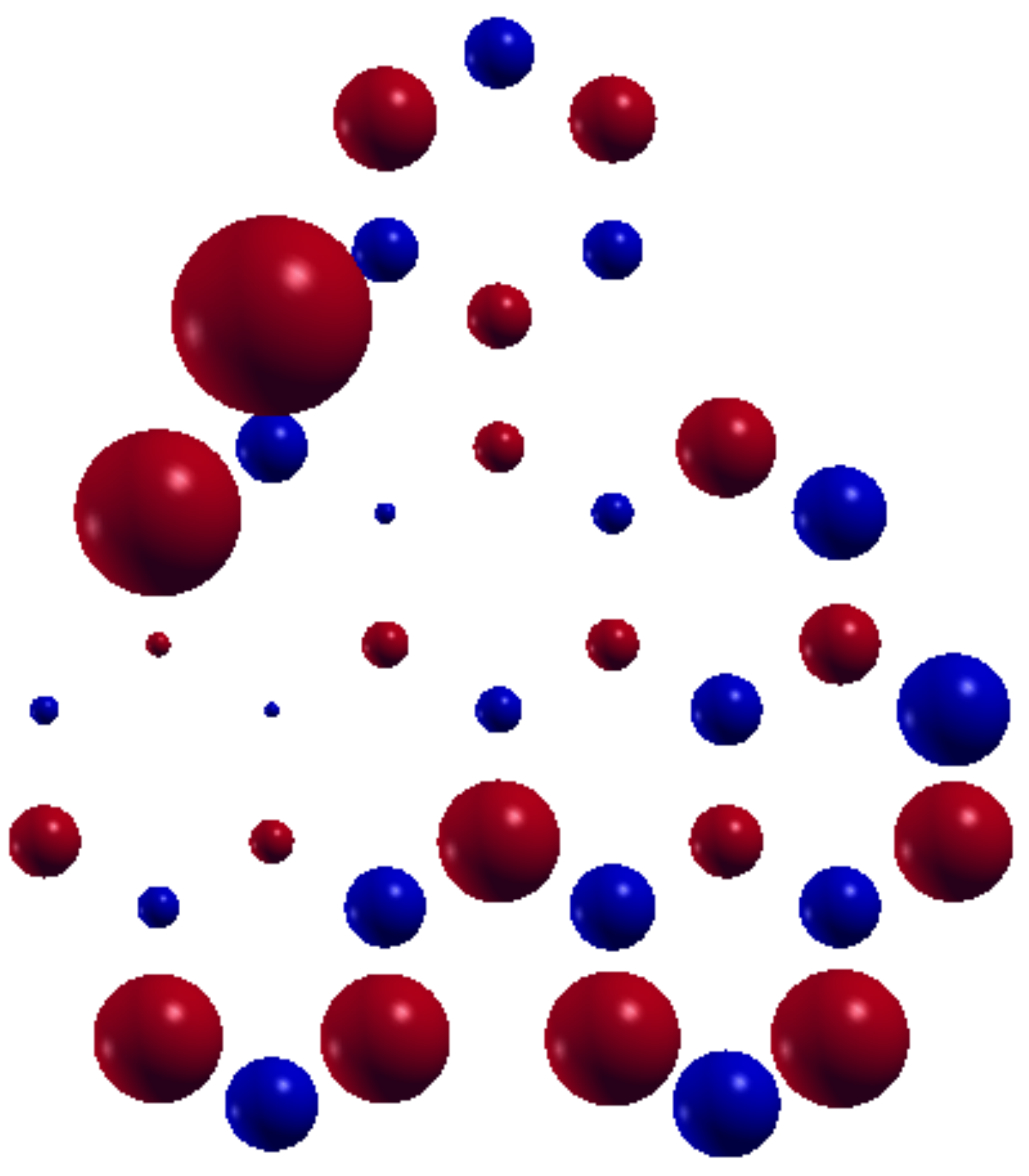}

}\subfloat[GQD-40]{\includegraphics[width=1.4cm]{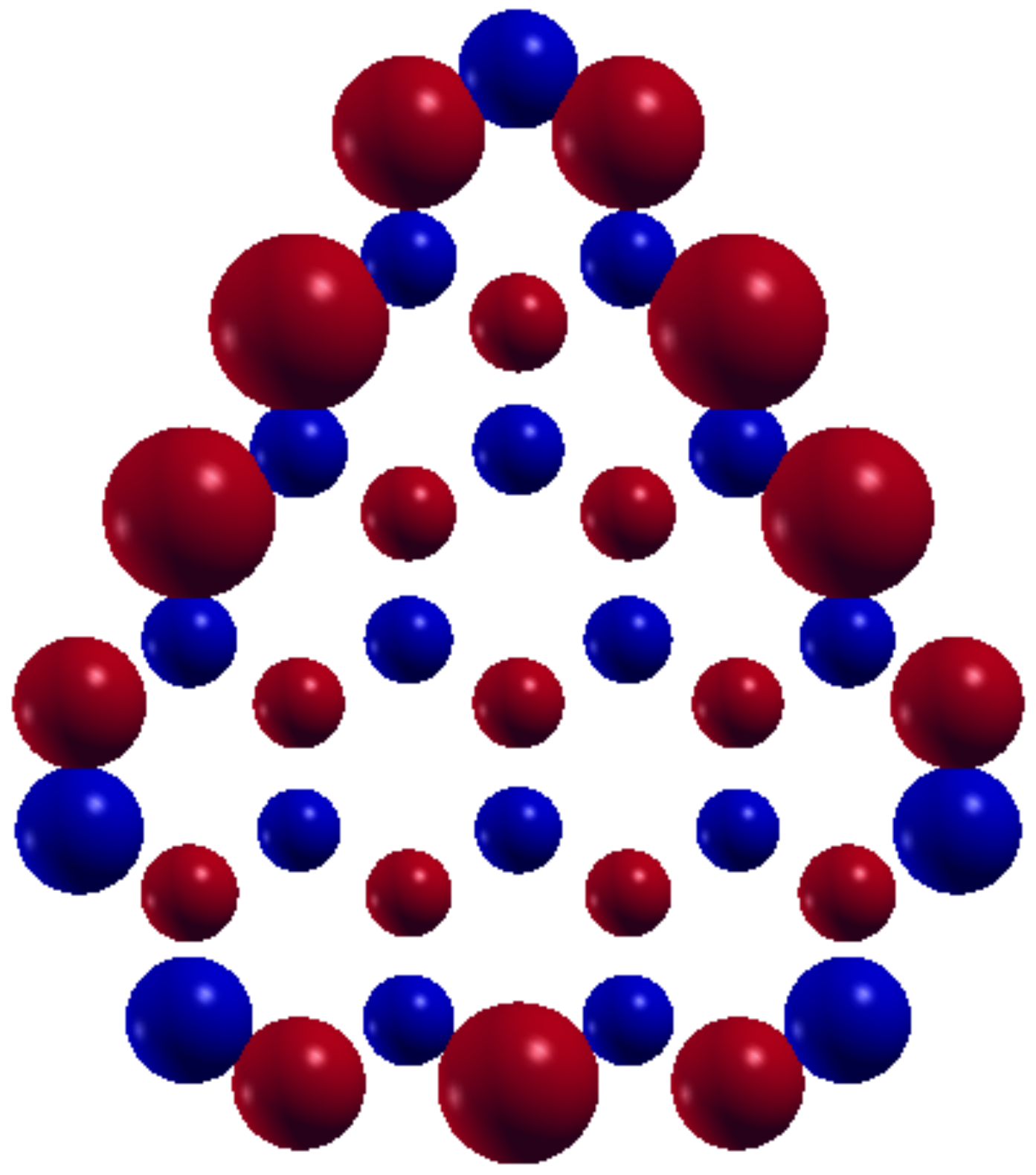}}\subfloat[GQD-48]{\includegraphics[width=1.4cm]{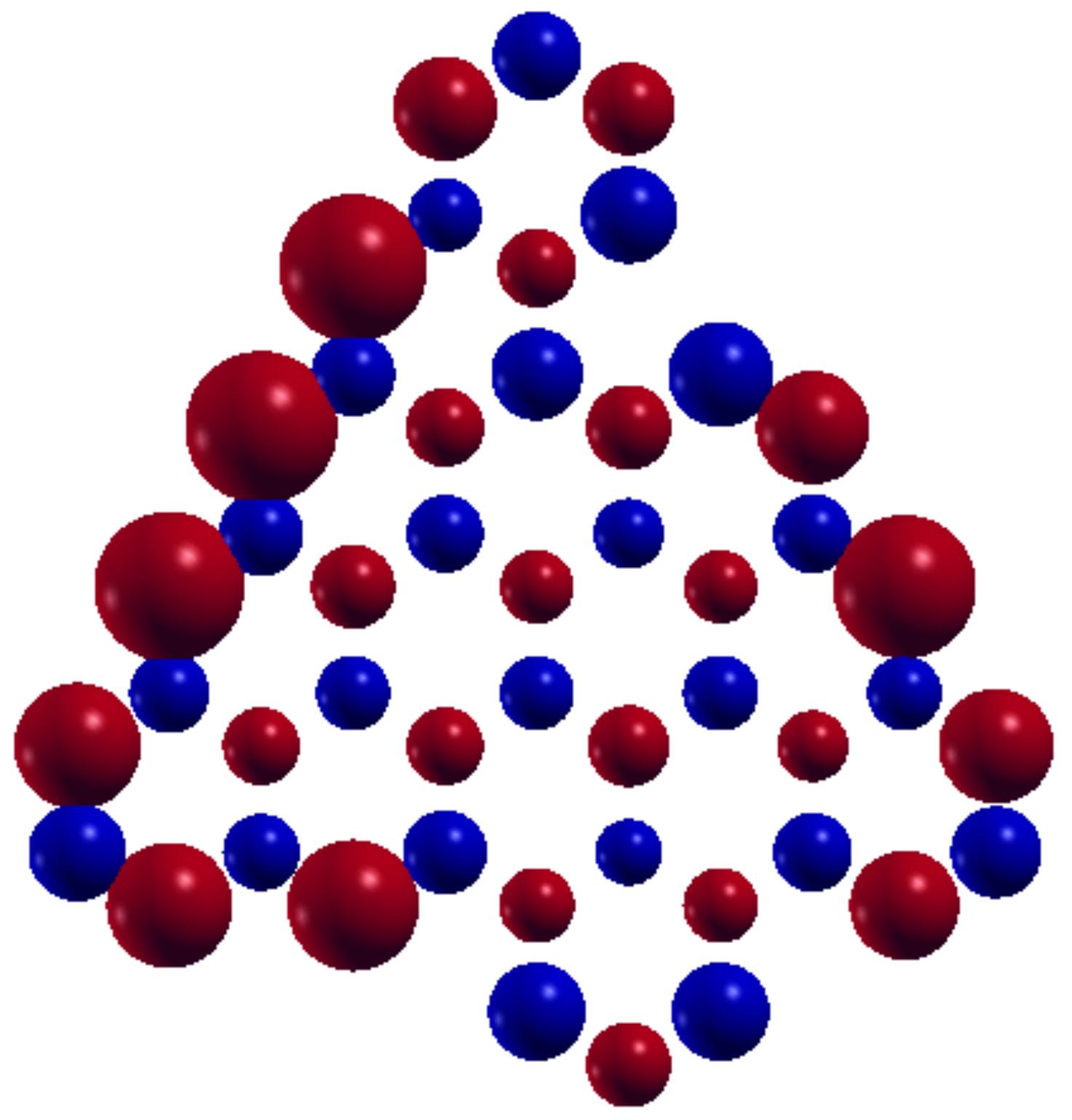}}\caption{Spin-density plots of FM \label{fig:Spin-density-plots-of-FM}state
of: (a) BQD-38, (b) RQD-54, (c) GQD-38, (d) GQD-40, and (e) GQD-48,
in the absence of electric field. The red and the blue spheres represent
the two different spin orientations (up/down or $\alpha/\beta$) of
the carbon atoms. }
\end{figure}

Figure \ref{fig:Spin-density-plots-of-FM} represents spin-density
plots of FM state of BQD-38, RQD-54, GQD-38, GQD-40 and GQD-48, in
the absence of electric field. It is observed that the spin density
is largest on the zigzag edges and decreases rapidly from the zigzag
edge to the middle of the QD. In addition, the spin density corresponding
to (up ($\alpha$) taken as majority spin) spin direction is more
as compared to that of the other spin orientation giving rise to the
FM character. In case of $D_{2h}$ symmetry QDs (RQD-54 and BQD-38),
the spin density is uniform at the opposite edges of the quantum dots.
However, in case of lower symmetry (GQD-40) or completely asymmetric
QDs (GQD-38 and GQD-48), the spin-density corresponding to majority
spin orientation (up or $\alpha$) is concentrated more on one side
of the dot, as compared to the other side.

\subsection{External electric field driven magnetic phase transitions of graphene
quantum dots }

\begin{figure}[h]
\subfloat{\includegraphics[width=4cm]{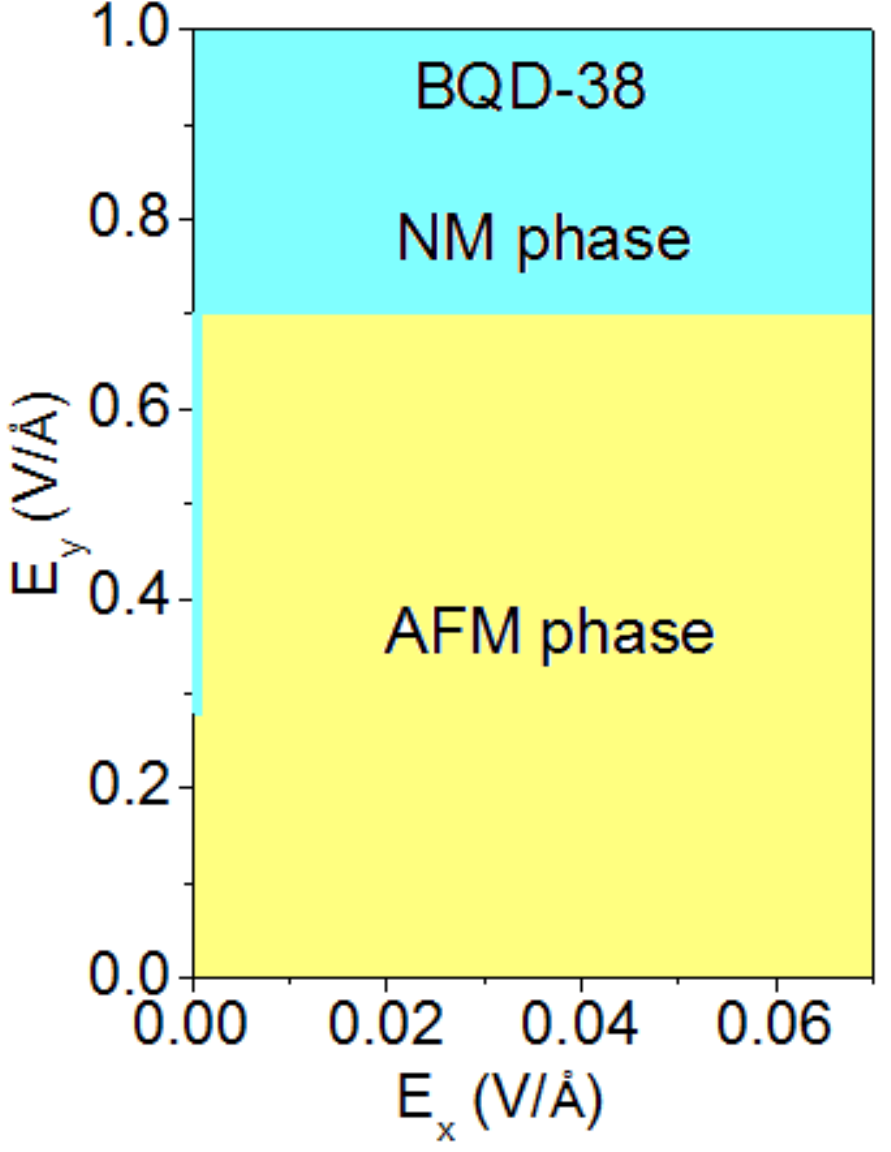}

}\subfloat{\includegraphics[width=4cm]{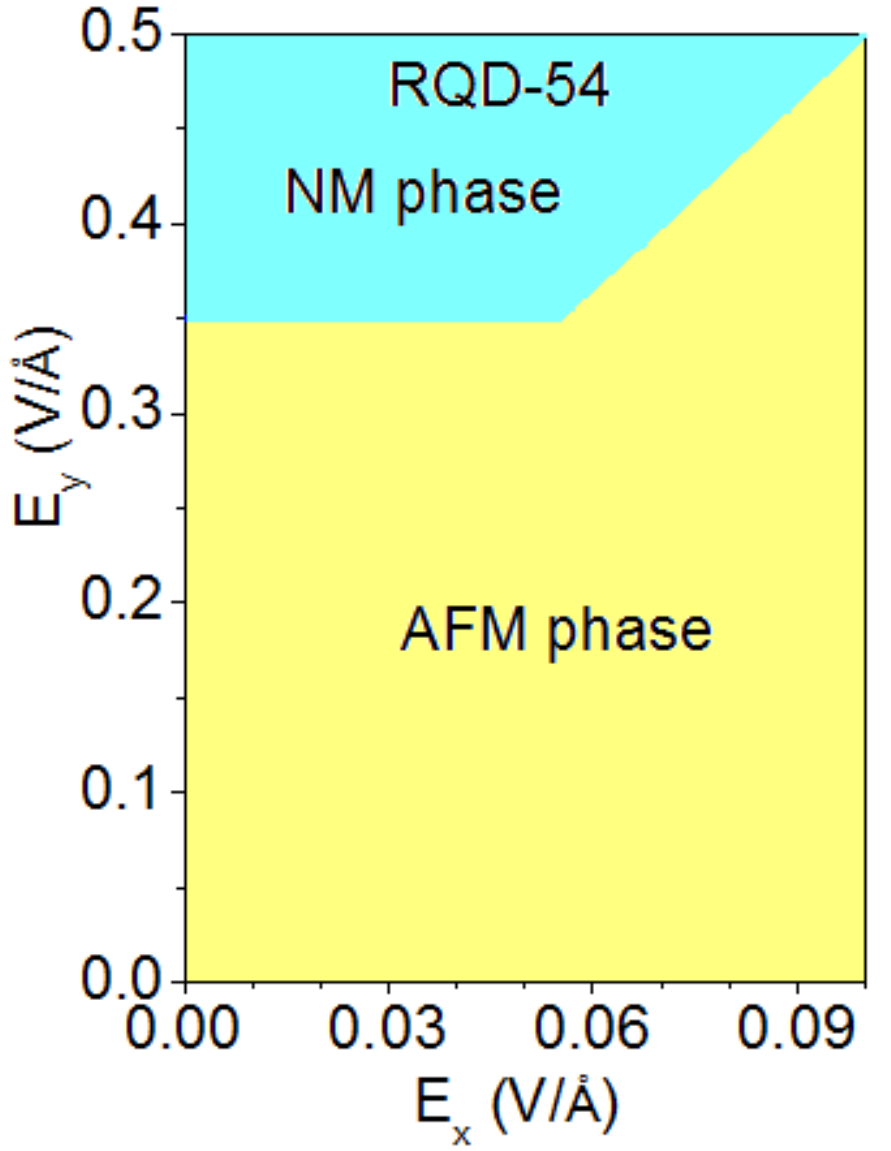}

}

\subfloat{\includegraphics[width=4cm]{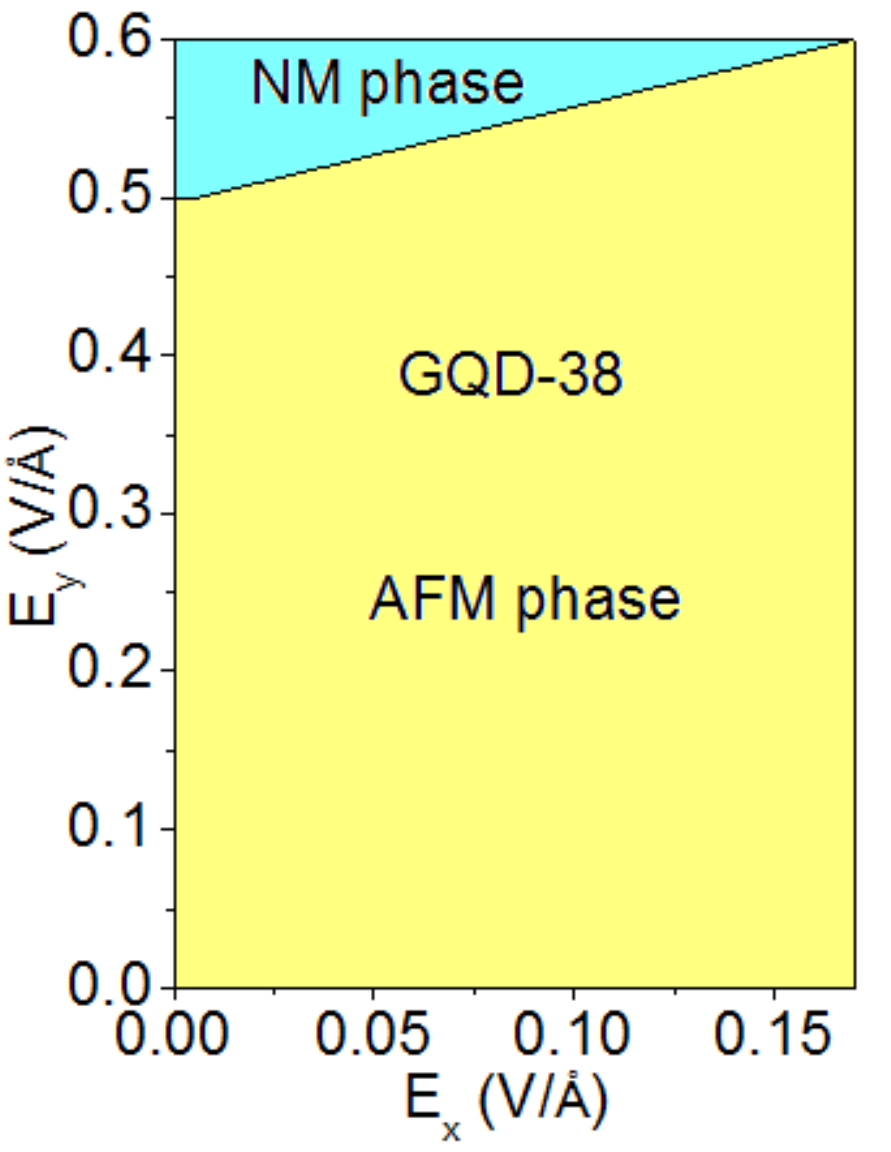}}\subfloat{\includegraphics[width=4cm]{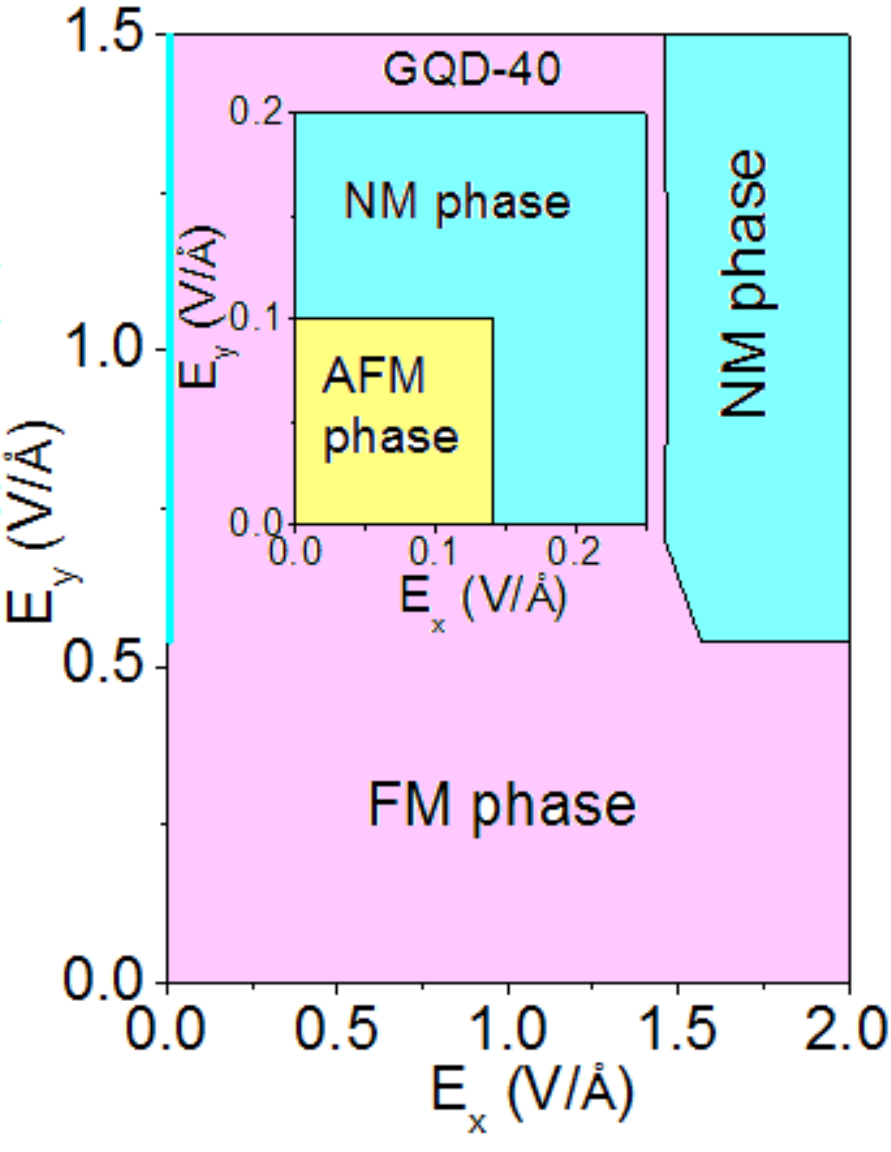}

}

\subfloat{\includegraphics[width=4cm]{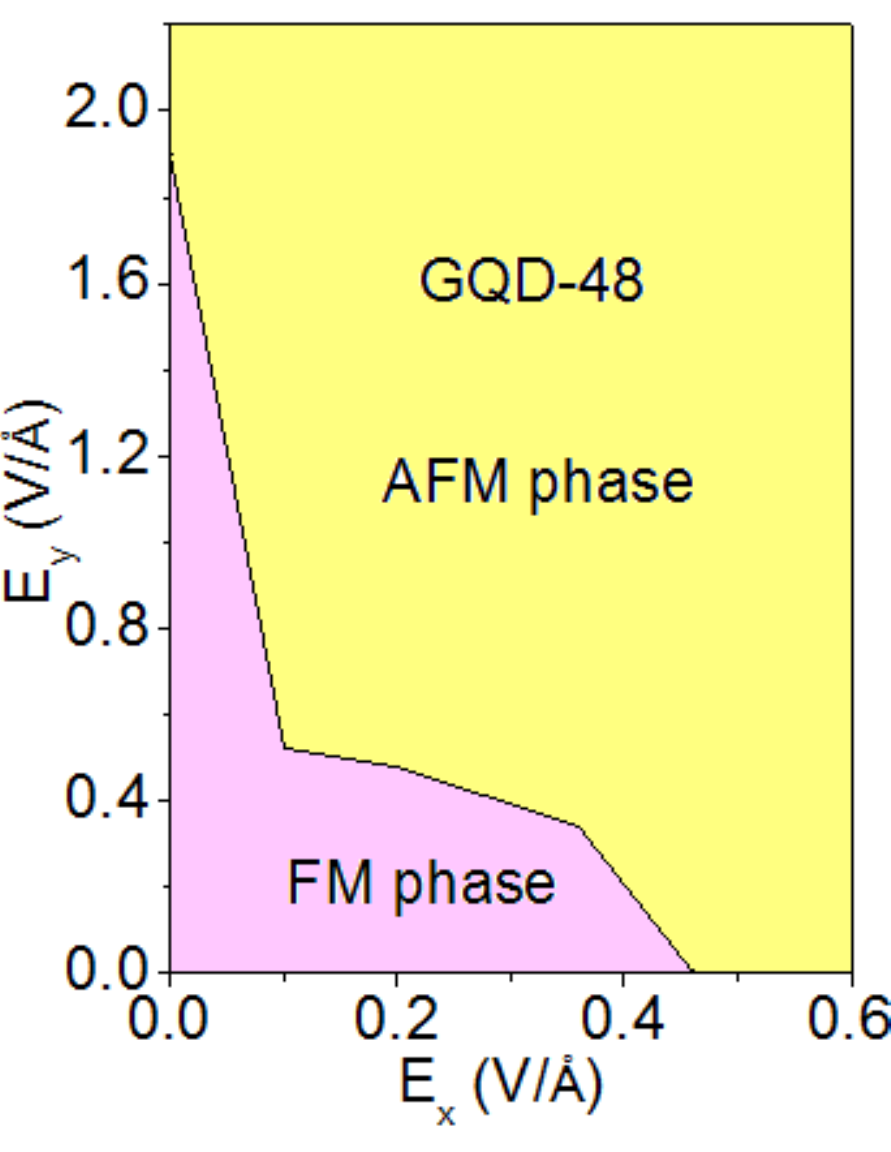}

}\caption{Phase diagrams \label{fig:Phase-diagrams-of}of the different magnetic
phases of BQD-38, RQD-54, GQD-38, GQD-40 and GQD-48 in the presence
of electric field. The inset of GQD-40 represents a phase transition
from the first excited state (AFM) to the second excited state (NM).}
\end{figure}

Figure \ref{fig:Phase-diagrams-of} presents the magnetic phase diagrams
of BQD-38, RQD-54, GQD-38, GQD-40 and GQD-48, with the external electric
field in the plane of the QDs being the control parameter. It is observed
that there is no phase transition for QDs with balanced sublattices
(BQD-38, RQD-54, GQD-38), under the influence of a longitudinal electric
field ($E_{x}$). However, when these QDs are exposed to a transverse
electric field ($E_{y}$), AFM order gets destroyed, resulting in
an NM state. The intermediate FM phase for RQD-54 and BQD-38 is not
achieved during the transition from AFM to NM state. Further, BQD-38,
RQD-54, and GQD-38, undergo a phase change from AFM to NM configuration
when subjected to an electric field in the $xy$ plane, with unequal
$x/y$ components. Initially the NM phase of BQD-38 is stable when
exposed to $E_{y}$($\geq$0.28 V/Å ) but disappears at the instant
$E_{x}$ is switched on. However, at high values of $E_{y}$ ($\geq$
0.7 V/Å), the NM state is energetically stable under the influence
of both $E_{x}$ and $E_{y}$. In case of GQD-40, the ground state
(FM) remains stable, while a phase transition occurs from the first
excited state (AFM) to the second excited state (NM), when a low strength
electric field is applied in any direction in the $xy$ plane, as
depicted in the inset of the phase diagram of GQD-40. 

The ground state (FM) remains unchanged when GQD-40 is exposed to
a purely longitudinal field. However, at higher values of $E_{y}$
($\geq$0.54 V/Å ), a phase change from the ground state (FM) to NM
configuration is observed. For this magnitude of $E_{y}$, the NM
configuration is destroyed, at the instant $E_{x}$ is also applied.
However, the NM phase becomes energetically stable at this value of
$E_{y}$, for higher strength of $E_{x}$ ($\geq$1.57 V/Å ). The
ground state (FM) of GQD-48, on the other hand, undergoes a phase
transformation to the AFM phase, with the application of an electric
field in any direction in the $xy$ plane. In this case, in contrast
to GQD-40, an initial phase change between the excited states is not
observed. Thus, direction of electric field plays an important role
in the tuning of phase transitions exhibited by QDs.

\subsection{Dependence of spin-polarized band-gaps of AFM and FM phases on external
electric field}

\begin{figure}
\subfloat{\includegraphics[scale=0.3]{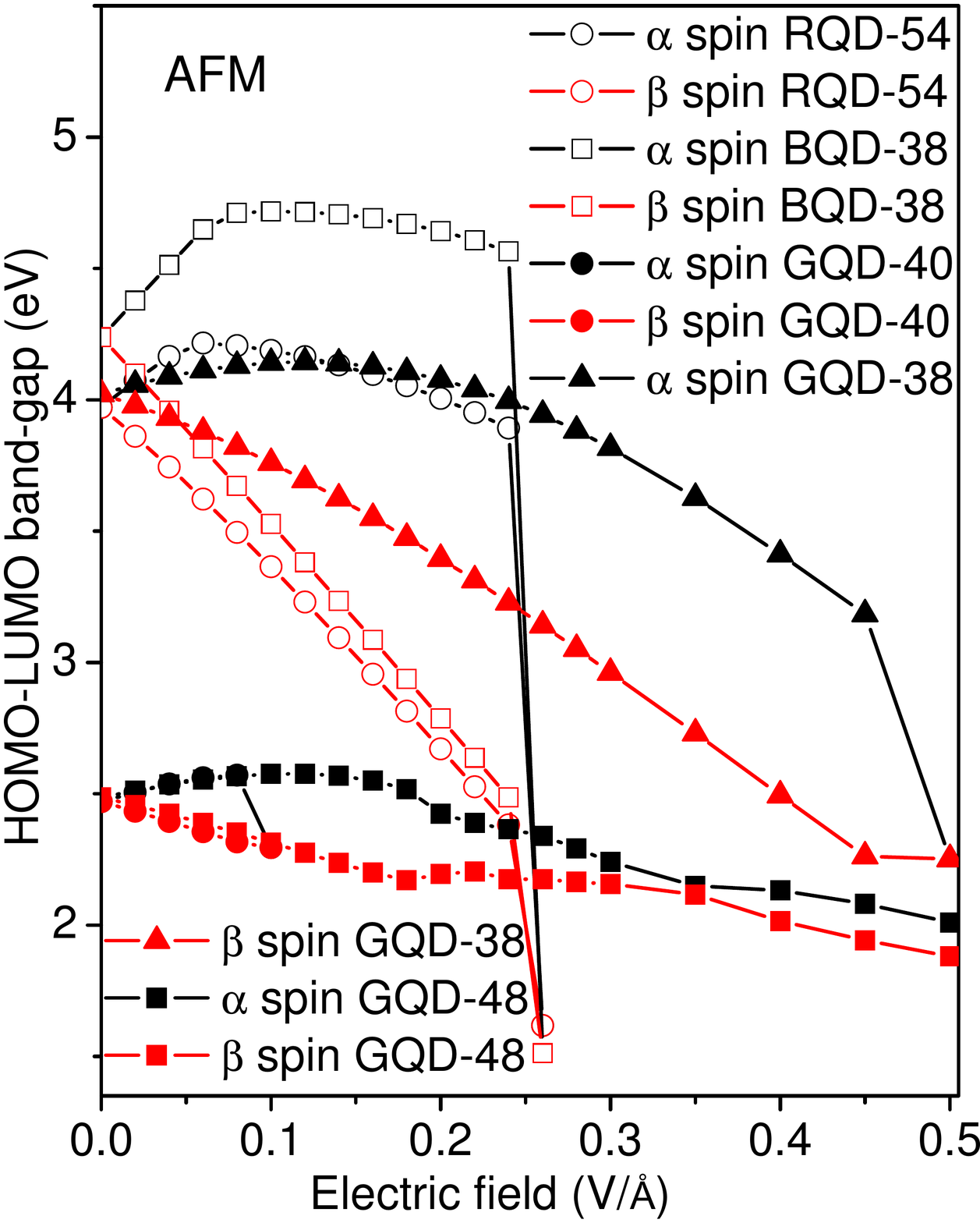}}

\subfloat{\includegraphics[scale=0.3]{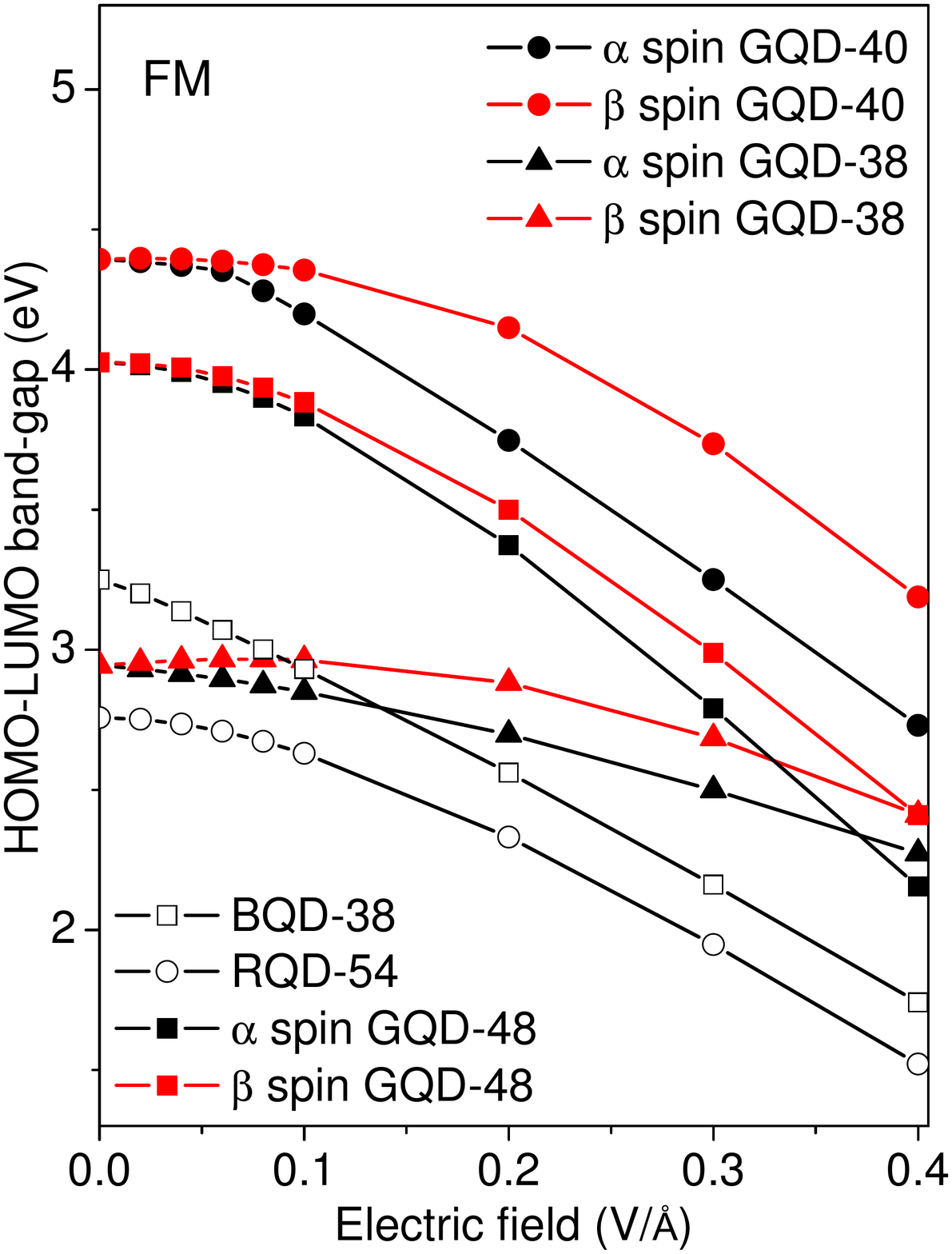}

}\caption{Variation of \label{fig:Variation-of-optical}spin-polarized HOMO-LUMO
band-gap of AFM and FM phase of QDs as a function of an in-plane transverse
electric field.}
\end{figure}

Figure \ref{fig:Variation-of-optical} represents the variation in
the spin-polarized H-L band-gaps of AFM and FM phases of QDs, as functions
of transverse electric field. The H-L band-gap corresponding to the
two different spin-orientations ($\alpha$ and $\beta$) are degenerate
in the absence of electric field, for both the AFM, and the FM configurations.
In case of the AFM phase, the application of electric field results
in a splitting of the band-gaps for up and down spins. The band-gaps
of spin-down ($\beta$) electrons decrease uniformly, while those
of up-spin ($\alpha$) electrons increases. However, the band-gap
of $\beta$ electrons never closes due to finite-size effect of the
quantum dots.\citep{ZhengdopedPhysRevB.78.155118} Thus, the behavior
of the spin-polarized band-gap under the influence of an external
electric field is not exactly half-metallic. With the increasing electric
field, the band-gap splitting decreases, and eventually the gaps for
$\alpha$ and $\beta$ spins again become degenerate for all the QDs
considered, except GQD-48. The band-gap splitting in AFM state arises
due to the spatial localization of the spin densities corresponding
to $\alpha$ and $\beta$ spins at the opposite edges of the quantum
dot (Fig. \ref{fig:Spin-density-plots-of-AFM}).  The application
of electric field leads to a spin transfer, due to charge transfer,
between the opposite corners of QDs, resulting in a decrease in spatial
localization of spins, and the band-gap splitting. The degeneracy
of the spin-polarized band gaps at higher electric fields is due to
phase transition of AFM state to NM state. Sharp drops in the band
gaps of RQD-54 and BQD-38 leading to the spin degeneracy are due to
the large difference between the band gaps of the AFM and NM states
(Table \ref{tab:HOMO-LUMO-band-gap-and-1}) of these QDs. This splitting
of the H-L band-gap is always evident when the QDs are subjected to
an in-plane transverse or diagonal electric field, while it is absent
for an in-plane longitudinal electric field for BQD-38, RQD-54, GQD-38,
and GQD-40. The band-gap splitting is more pronounced for QDs (RQD-54,
BQD-38) with high symmetry ($D_{2h}$ symmetry), as compared to QDs
(GQD-38, GQD-40 and GQD-48) which exhibit either a lower symmetry
($C_{2v}$ symmetry), or are completely asymmetric. 

For FM configurations, the H-L band-gap corresponding to $\alpha$
and $\beta$ spins remains degenerate under the influence of electric
field in any direction, for $D_{2h}$ symmetry QDs (RQD-54 and BQD-38),
and it decreases with the increasing field. However, for lower symmetry
(GQD-40 with $C_{2v}$ symmetry) or asymmetric QDs (GQD-38 and GQD-48),
the electric field splits the gap for the two spin orientations. With
the increasing field strength, the gap corresponding to one spin orientation
($\alpha$) decreases more rapidly as compared to that of the other
spin ($\beta$). This H-L gap splitting always occurs in the presence
of in-plane transverse and diagonal electric fields, while it is absent
for longitudinal electric field for GQD-40. This band-gap splitting
in the FM phase is due to the greater concentration of spin-density
of the same spin flavour at one side, as compared to the other side
of the quantum dot (Fig. \ref{fig:Spin-density-plots-of-FM}). Further,
in contrast to the AFM phase, the gaps corresponding to the $\alpha$
and $\beta$ spins in the FM state never become degenerate, even for
extremely high values of the electric field. This unique property
of tuning of the spin-dependent band-gap of AFM and FM arrangements
of QDs by electric fields of different strengths and alignments can
be effectively used in the field of spintronics. 

\subsection{Electroabsorption spectra of various magnetic phases of QDs }

Next, we analyze the EA spectra, i.e., optical absorption spectra
as a function of external electric field, of various magnetic phases
of QDs. The salient features exhibited by the EA spectra of the AFM
and FM phases of BQD-38, RQD-54, GQD-40, GQD-48 and GQD-38 indicate
that, in the absence of a magnetic phase diagram, the EA spectra is
self-sufficient to predict not only the magnetic ground state, but
also the states attained by the QDs after the phase transition. Next,
we discuss the calculated EA spectra of different classes of QDs considered
in this work. 

\subsubsection{Highly symmetric QDs with $D_{2h}$ symmetry }

\begin{figure}[h]
\includegraphics[scale=0.45]{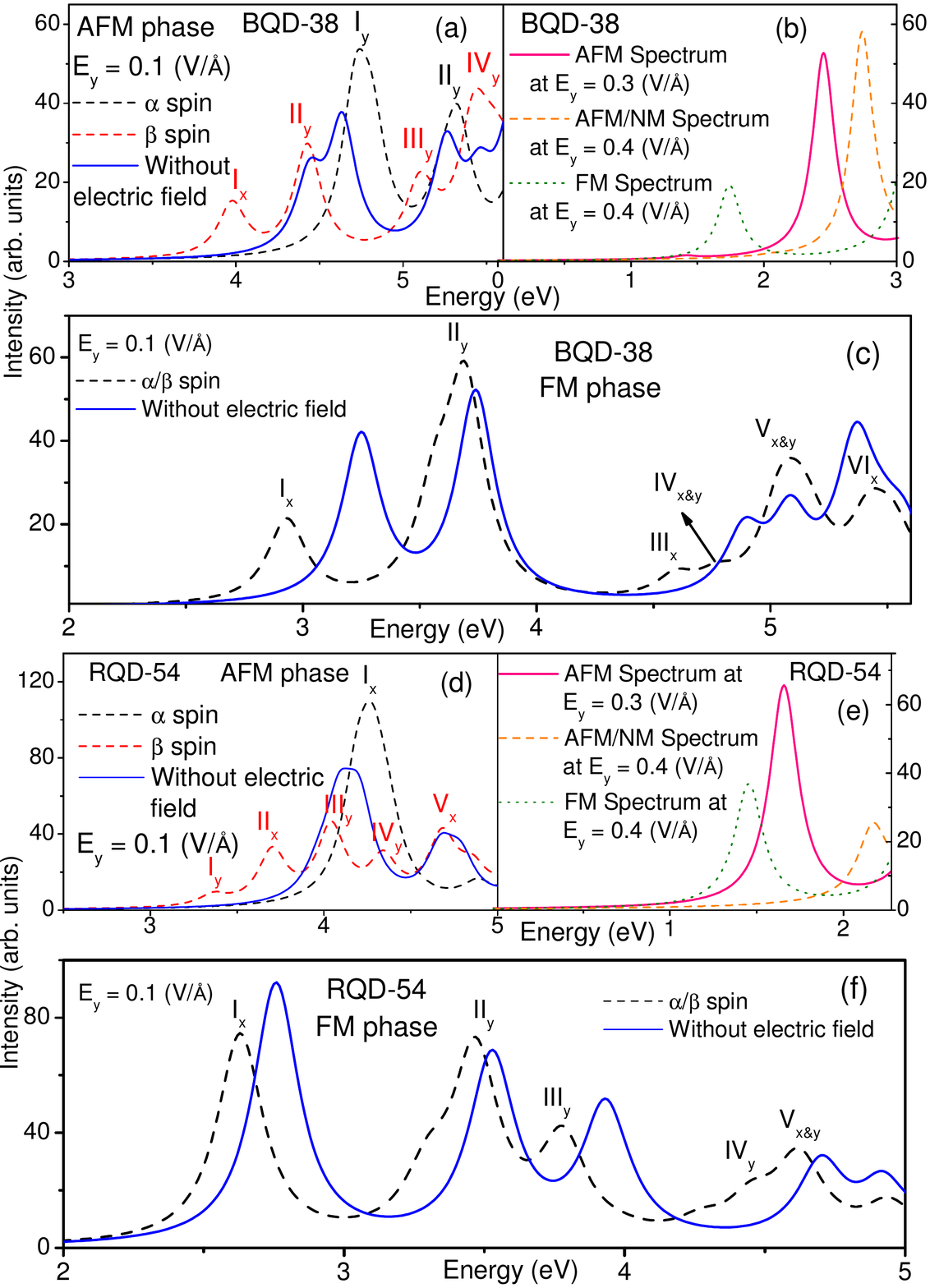}\caption{Computed \label{optic_symmetric-transverse}EA spectrum broadened
with a uniform line-width of 0.1 eV for the (a) AFM phase of BQD-38
(b) AFM (pink solid line) at $E_{y}$$=$0.3 V/$\textrm{\AA}$, NM
(orange dashed line) and FM phase (green dotted line) at $E_{y}$$=$0.4
V/$\textrm{\AA}$, of BQD-38 (c) FM phase of BQD-38 (d) AFM phase
of RQD-54 (e) AFM (pink solid line) at $E_{y}$$=$0.3 V/$\textrm{\AA}$,
NM (orange dashed line) and FM phase (green dotted line) at $E_{y}$$=$0.4
V/$\textrm{\AA}$, of RQD-54 and (f) FM phase of RQD-54. The red and
black dotted lines indicate the optical spectra for spin-down ($\beta$
spin) and spin-up ($\alpha$ spin) orbitals, respectively, in the
presence of E$_{y}$= 0.1 V/$\textrm{\AA}$. The blue solid line indicates
the optical spectrum in the absence of electric field.  Peak labels
imply peak numbers, with the subscripts indicating the polarization
directions.}
\end{figure}

\begin{figure}[h]
\includegraphics[scale=0.4]{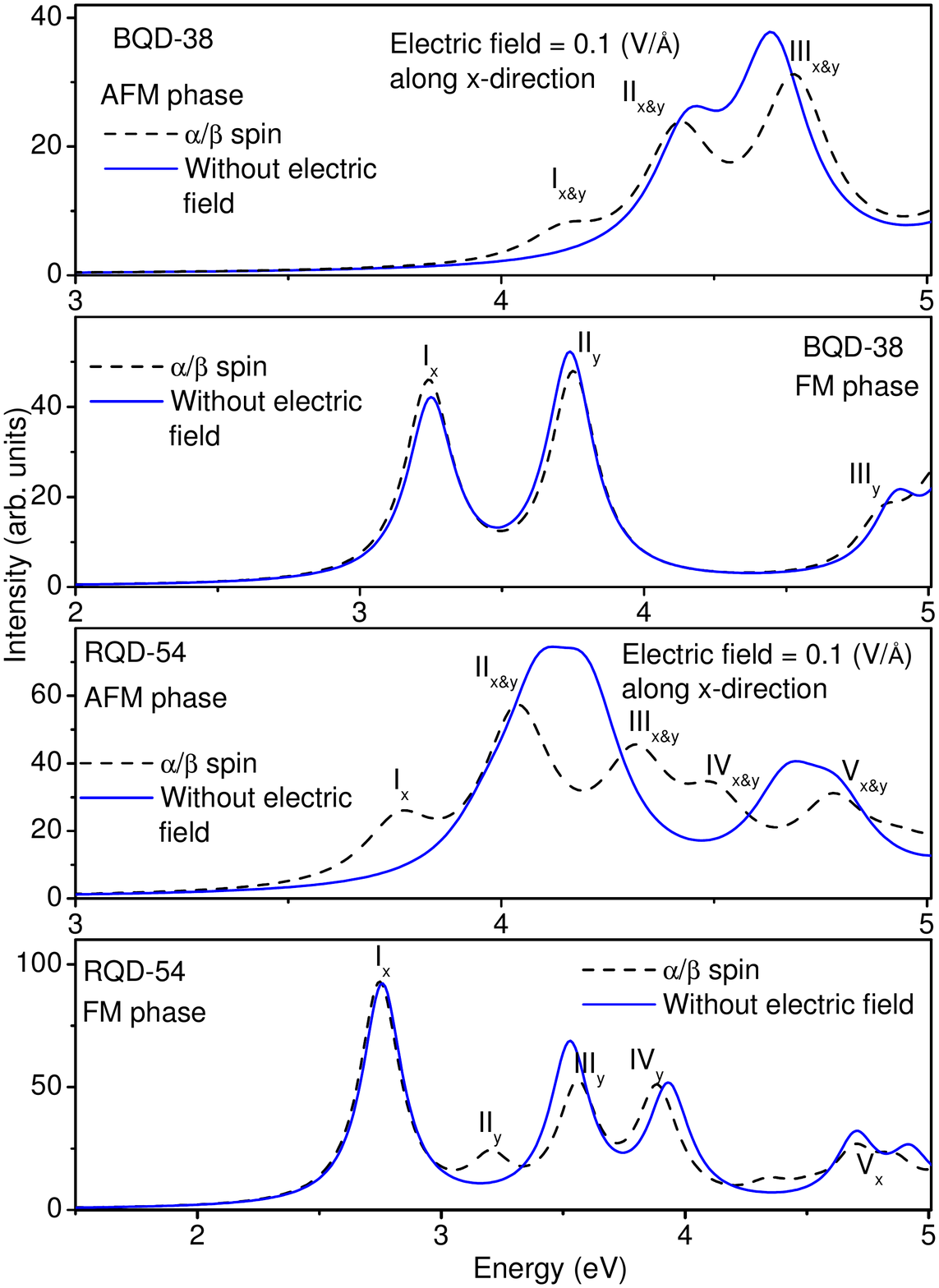}\caption{Computed \label{optic_symmetric-longitudinal}EA spectrum broadened
with a uniform line-width of 0.1 eV for the AFM and FM phases of BQD-38
and RQD-54. The black dotted line indicates the spectra for both the
spin-up ($\mbox{\ensuremath{\alpha}}$ spin) and spin-down ($\beta$
spin) electrons, in the presence of an in-plane longitudinal electric
field of 0.1 V/$\textrm{\AA}$. The blue solid line indicates the
absorption spectrum in the absence of electric field.  Peak labels
imply peak numbers, with the subscripts indicating the polarization
directions. }
\end{figure}

\begin{figure}[h]
\includegraphics[scale=0.4]{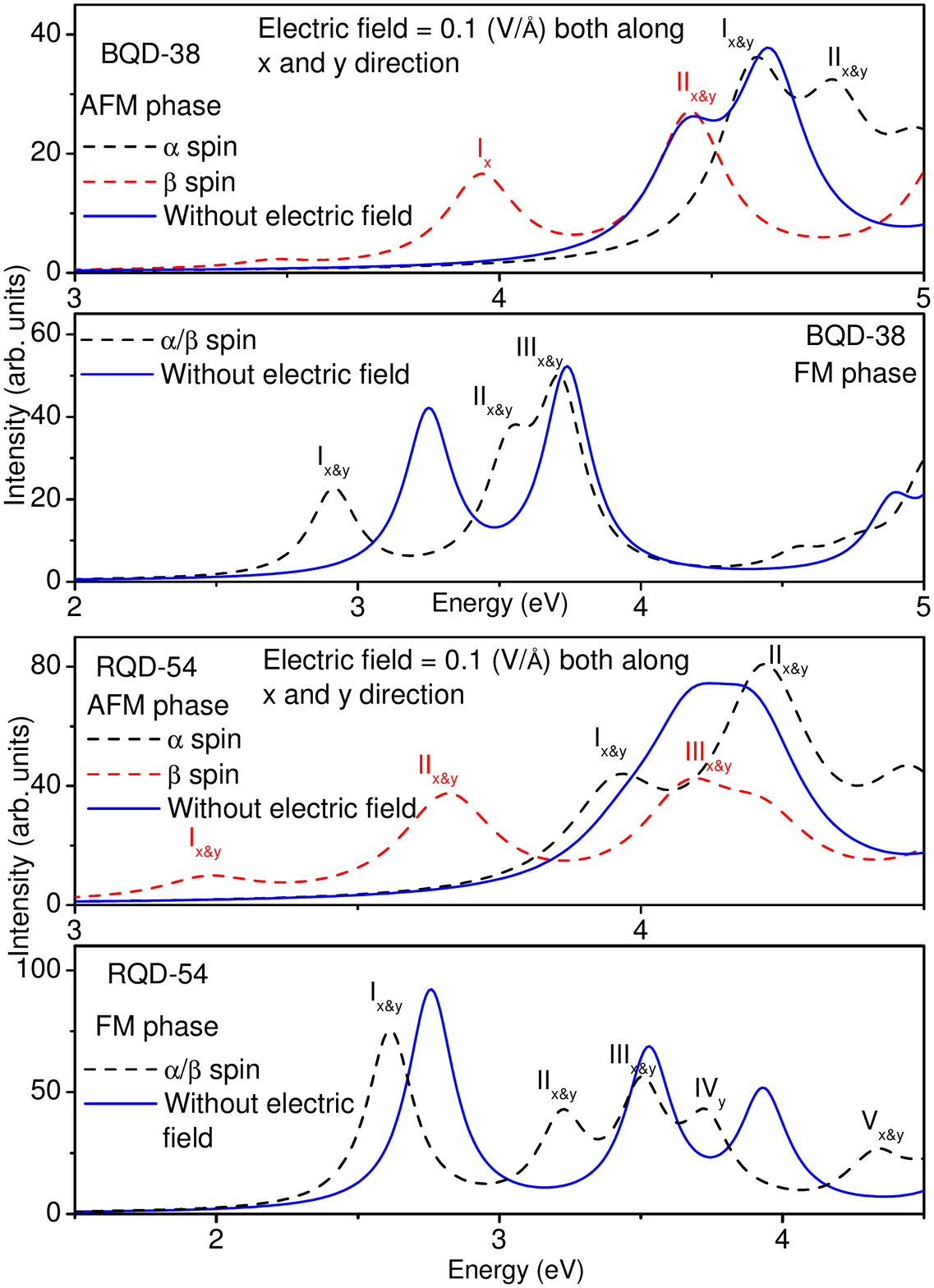}\caption{Computed \label{optic_symmetric-diagonal}EA spectrum broadened with
a uniform line-width of 0.1 eV for the AFM and FM phases of BQD-38
and RQD-54. The red and black dotted lines indicate the spectra for
spin-down ($\beta$ spin) and spin-up ($\alpha$ spin) electrons,
respectively, in the presence of an in-plane diagonal electric field
of 0.1 V/$\textrm{\AA}$. Peak labels imply peak numbers, with the
subscripts indicating the polarization directions. }
\end{figure}

The EA spectra of the AFM and FM phases of highly symmetric QDs with
$D_{2h}$ symmetry (BQD-38 and RQD-54) corresponding to transverse,
longitudinal and diagonal electric field, are presented in Figs. \ref{optic_symmetric-transverse},
\ref{optic_symmetric-longitudinal} and \ref{optic_symmetric-diagonal},
respectively. Salient features of the calculated absorption spectra
are as follows. 

\begin{enumerate}
\item The EA spectrum of AFM phase of BQD-38 and RQD-54 exhibits a spin-sensitive
split on applying transverse electric fields. (Fig. \ref{optic_symmetric-transverse}(a)
and Fig. \ref{optic_symmetric-transverse}(d)). However, this spin
splitting of the EA spectrum is not observed, in the presence of transverse
electric field, for the FM phases of BQD-38 (\ref{optic_symmetric-transverse}
(c)) and RQD-54 ( Fig. \ref{optic_symmetric-transverse} (f)). Thus,
this distinctive feature of the EA spectrum, in the presence of transverse
electric field, can be used to identify the ground state magnetic
configuration of these QDs.
\item At higher values of transverse electric fields ($E_{y}$$\approx$0.3
V/$\textrm{\AA}$), the spin splitting of the EA spectrum of the AFM
phase of BQD-38 and RQD-54 vanishes (pink solid line in figs. \ref{optic_symmetric-transverse}
(b) and \ref{optic_symmetric-transverse} (e)), indicating a phase
transition. In order to identify the state (FM or NM) attained by
the QDs after this phase transition, energy shifts of the EA spectra
with increasing transverse electric field ($E_{y}$$>$0.3 V/$\textrm{\AA}$)
are analyzed. If the AFM phase undergoes a phase change to NM state
at $E_{y}$$\approx$0.3 V/$\textrm{\AA}$, the EA spectrum gets blue
shifted at electric fields $E_{y}$$>$0.3 V/$\textrm{\AA}$ (see
orange dashed lines in Figs. \ref{optic_symmetric-transverse} (b)
and \ref{optic_symmetric-transverse} (e) at $E_{y}$$=$0.4 V/$\textrm{\AA}$).
On the other hand, if the FM state is reached after the phase transition,
the EA spectrum gets red shifted at electric fields $E_{y}$$>$0.3
V/$\textrm{\AA}$ (see green dotted lines in Figs. \ref{optic_symmetric-transverse}
(b) and \ref{optic_symmetric-transverse} (e), at $E_{y}$$=$0.4
V/$\textrm{\AA}$). Since, the EA spectrum of the AFM configuration
at $E_{y}$$=$0.4 V/$\textrm{\AA}$ exhibits a blue-shift and coincides
with the EA spectrum of the NM state at $E_{y}$$=$0.4 V/$\textrm{\AA}$(figs.
\ref{optic_symmetric-transverse} (b) and \ref{optic_symmetric-transverse}
(e) for BQD-38 and RQD-54, respectively), it implies that the AFM
phase undergoes a phase transition to the NM state, as also predicted
by their phase diagrams (Fig. \ref{fig:Phase-diagrams-of}). 
\item The EA spectra of the AFM phases of BQD-38 and RQD-54\textcolor{red}{{}
}do not exhibit a spin-sensitive split on the application of longitudinal
electric fields (Fig. \ref{optic_symmetric-longitudinal}). However,
these spectra do exhibit split corresponding to opposite spin orientations,
when perturbed by an in-plane diagonal electric field (Fig. \ref{optic_symmetric-diagonal}).
Thus, we conclude that the presence of a transverse component in the
external electric field is essential for spin splitting to be observed
in the AFM phase EA spectra of these QDs.
\item In case of the FM states of BQD-38 and RQD-54, the EA spectra do not
exhibit any spin-sensitive split on application of in-plane longitudinal
and diagonal electric fields. Thus, spin-sensitive splitting of the
optical spectrum, on the application of electric field in any direction,
is not possible for the FM phase of BQD-38 and RQD-54.
\end{enumerate}

\subsubsection{Low symmetry QDs }

\begin{figure}[H]
\includegraphics[scale=0.35]{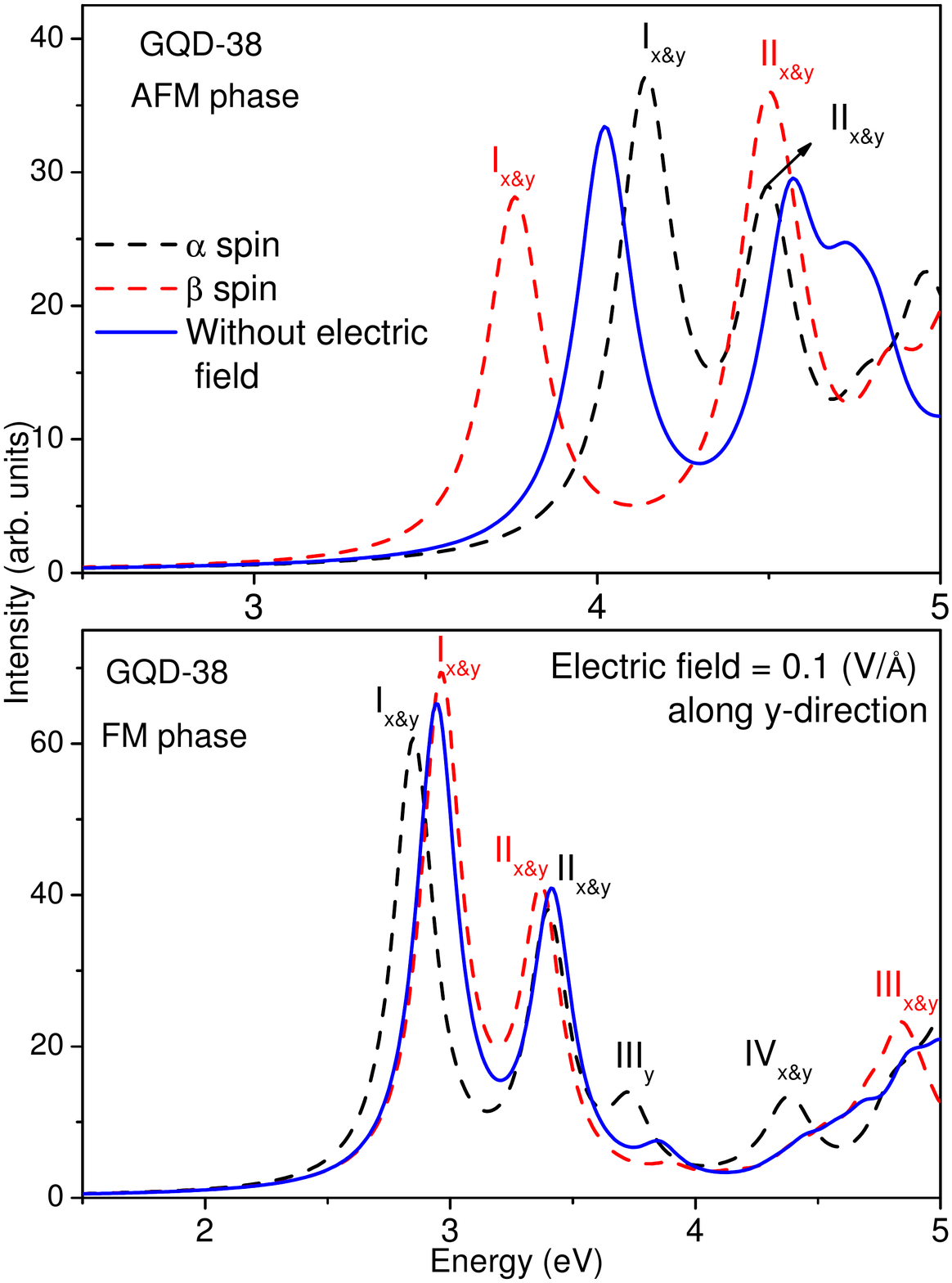}\caption{Computed EA spectrum \label{optic_asymm_unbal-transverse} broadened
with a uniform line-width of 0.1 eV for the AFM and FM phases of GQD-38.
The red and black dotted lines indicate the optical spectra for spin-down
($\beta$ spin) and spin-up ($\alpha$ spin) electrons, respectively,
in the presence of transverse electric field. The blue solid line
indicates the optical spectrum in the absence of electric field. Peak
labels imply peak numbers, with the subscripts indicating the polarization
directions.}
\end{figure}

\begin{figure}[h]

\includegraphics[scale=0.35]{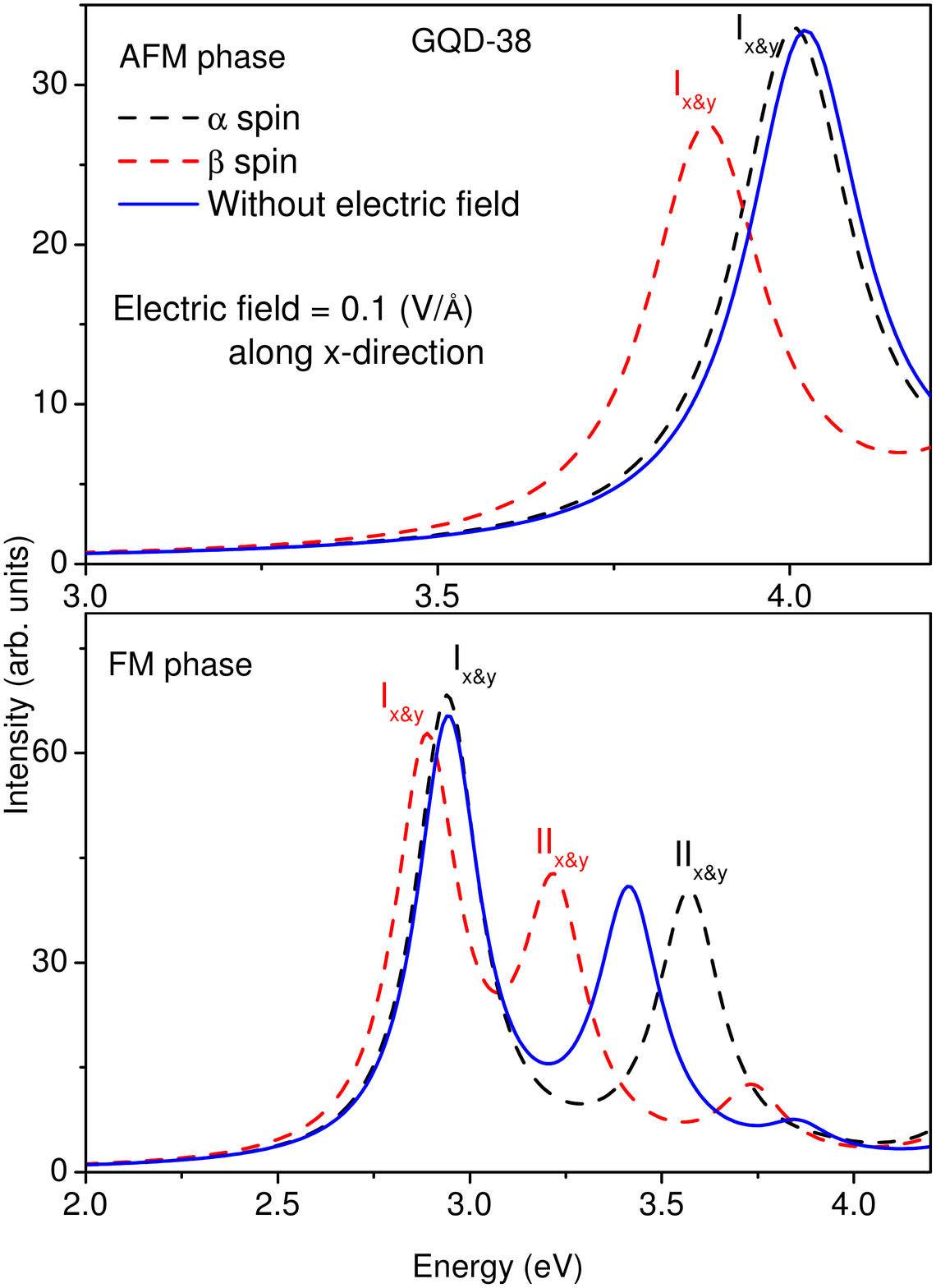}\caption{Computed EA spectrum \label{optic_asymm_unbal-longitudinal} broadened
with a uniform line-width of 0.1 eV for the AFM and FM phases of GQD-38.
The red and black dotted lines indicate the optical spectra for spin-down
($\beta$ spin) and spin-up ($\alpha$ spin) electrons, respectively,
in the presence of longitudinal electric field. The blue solid line
indicates the optical spectrum in the absence of electric field. Peak
labels imply peak numbers, with the subscripts indicating the polarization
directions.}

\end{figure}

\begin{figure}[H]

\includegraphics[scale=0.35]{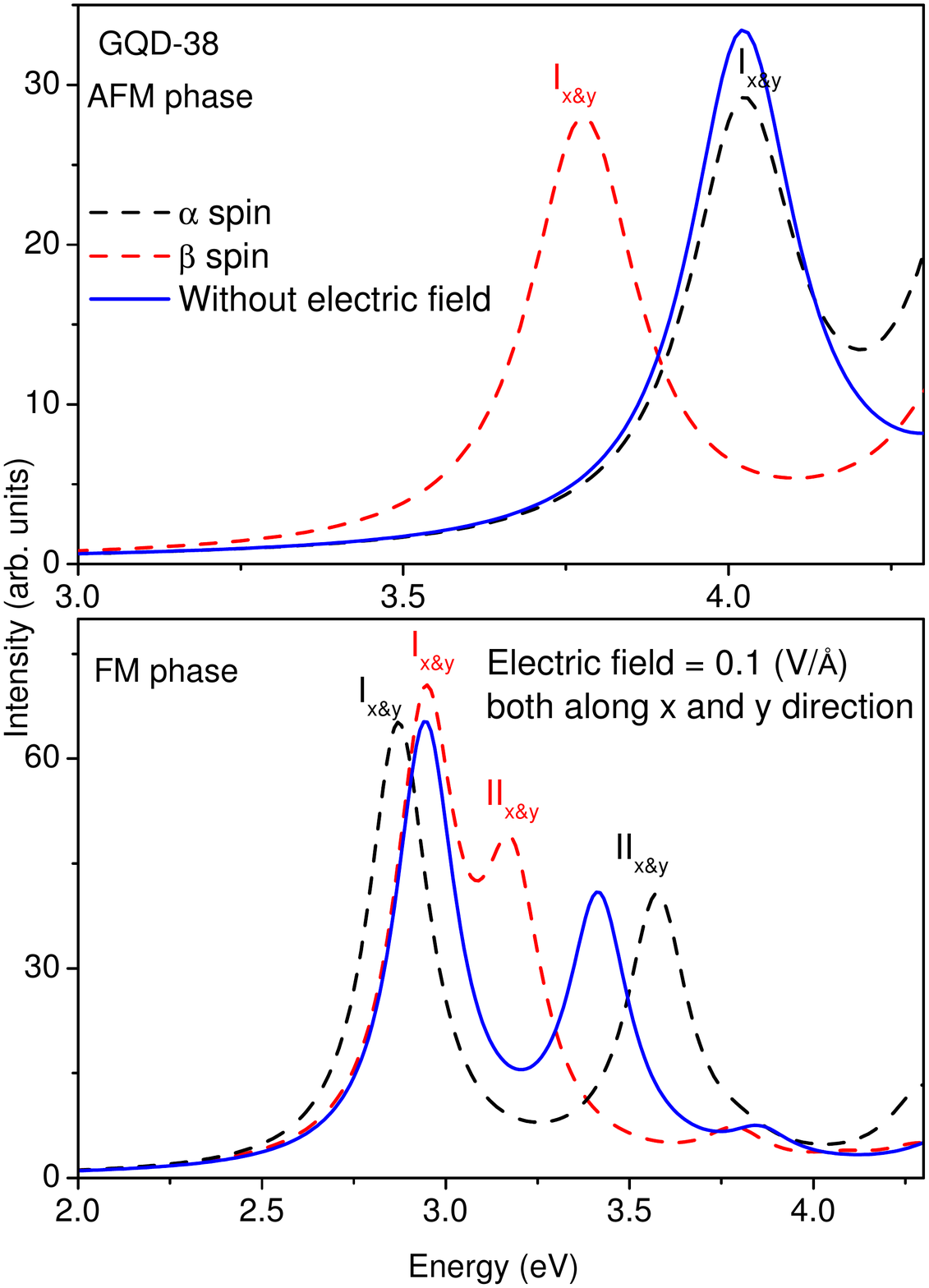}\caption{Computed EA spectrum \label{optic_asymm_unbal-diagonal} broadened
with a uniform line-width of 0.1 eV for the AFM and FM phases of GQD-38.
The red and black dotted lines indicate the optical spectra for spin-down
($\beta$ spin) and spin-up ($\alpha$ spin) electrons, respectively,
in the presence of a diagonal electric field. The blue solid line
indicates the optical spectrum in the absence of electric field. Peak
labels imply peak numbers, with the subscripts indicating the polarization
directions.}

\end{figure}

\begin{figure}[H]

\includegraphics[scale=0.4]{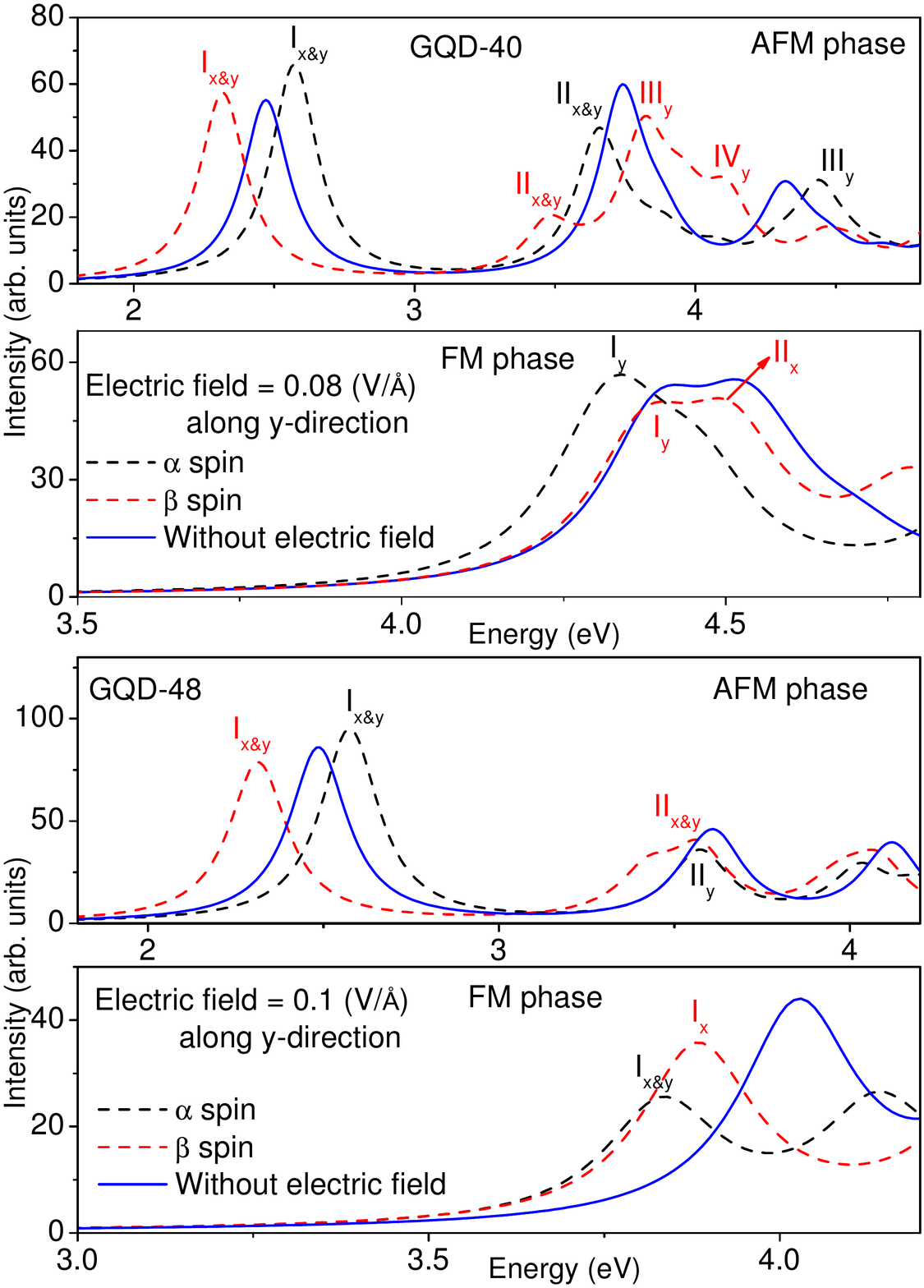}\caption{Computed EA spectrum \label{optic_asymm_GQD-40-GQD-48-transverse}
broadened with a uniform line-width of 0.1 eV for the AFM and FM phases
of GQD-40 and GQD-48. The red and black dotted lines indicate the
optical spectra for spin-down ($\beta$ spin) and spin-up ($\alpha$
spin) electrons, respectively, in the presence of a transverse electric
field. The blue solid line indicates the optical spectrum in the absence
of electric field. Peak labels imply peak numbers, with the subscripts
indicating the polarization directions.}

\end{figure}

\begin{figure}[H]

\includegraphics[scale=0.4]{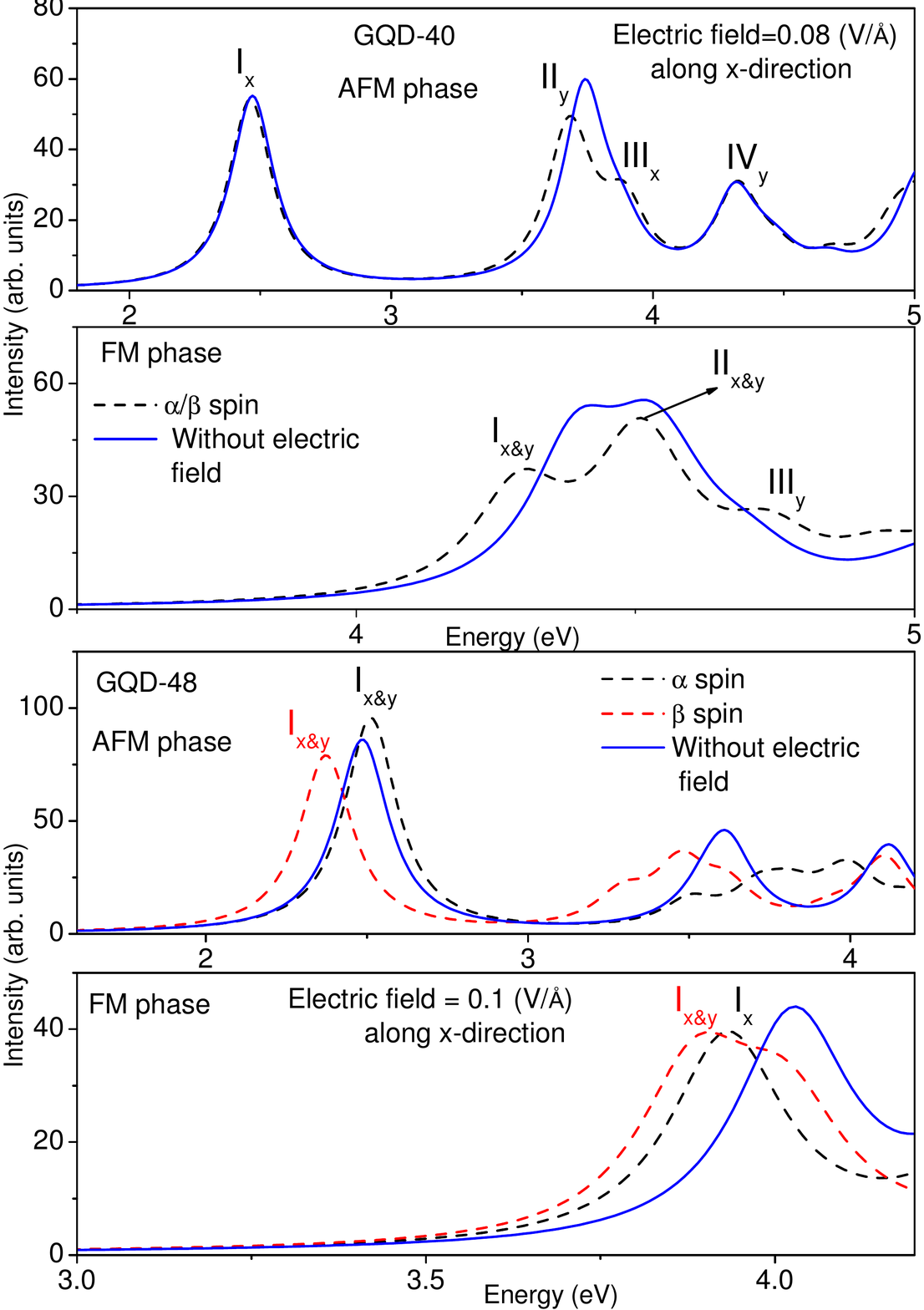}\caption{Computed EA spectrum \label{optic_asymm_GQD-40-GQD-48-longitudinal}
broadened with a uniform line-width of 0.1 eV for the AFM and FM phases
of GQD-40 and GQD-48. The red and black dotted lines indicate the
optical spectra for spin-down ($\beta$ spin) and spin-up ($\alpha$
spin) electrons, respectively, in the presence of a longitudinal electric
field. The blue solid line indicates the optical spectrum in the absence
of electric field. Peak labels imply peak numbers, with the subscripts
indicating the polarization directions.}

\end{figure}

\begin{figure}[H]

\includegraphics[scale=0.4]{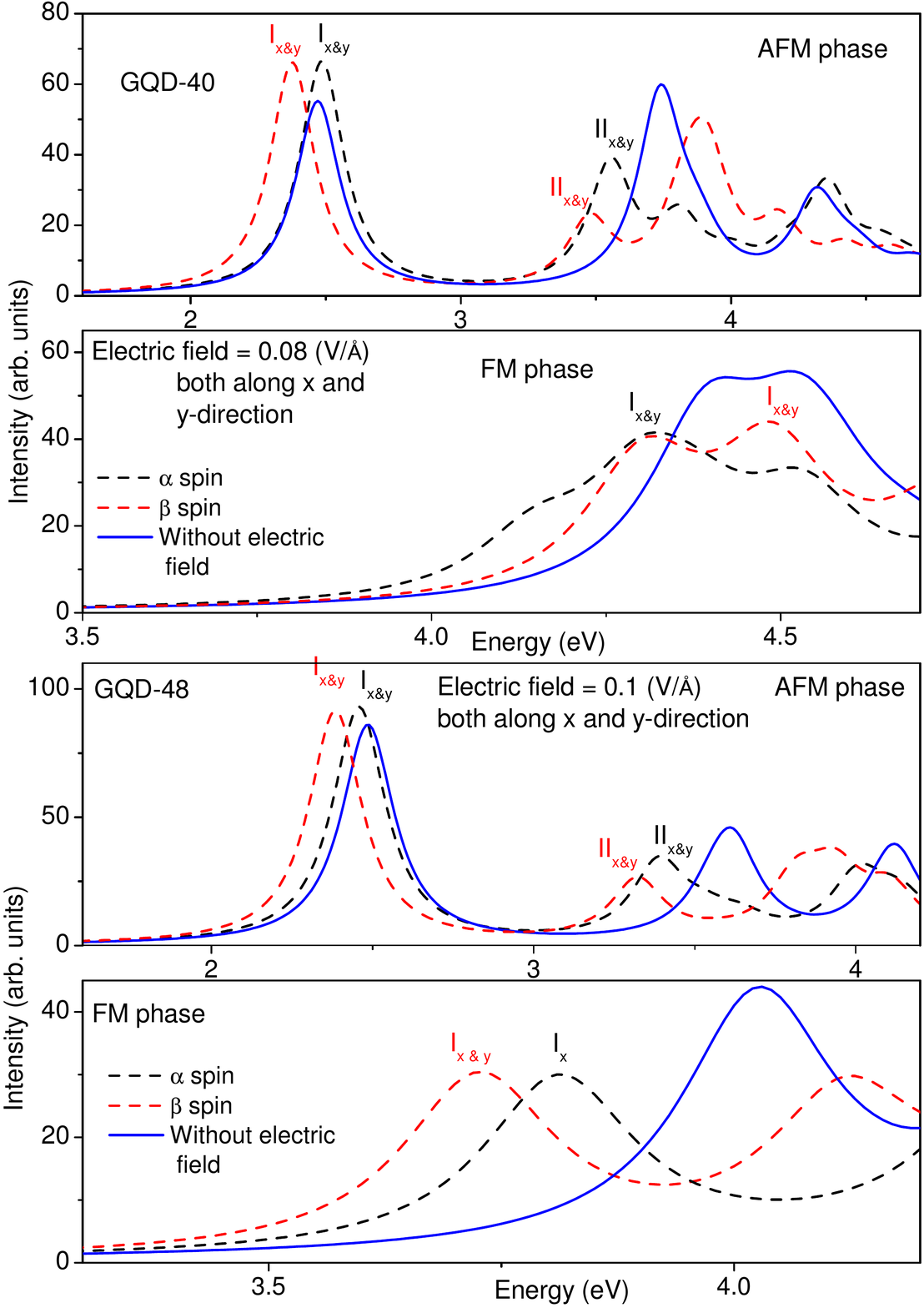}\caption{Computed EA spectrum \label{optic_asymm_GQD-40-GQD-48-diagonal} broadened
with a uniform line-width of 0.1 eV for the AFM and FM phases of GQD-40
and GQD-48. The red and black dotted lines indicate the optical spectra
for spin-down ($\beta$ spin) and spin-up ($\alpha$ spin) electrons,
respectively, in the presence of diagonal electric field. The blue
solid line indicates the optical spectrum in the absence of electric
field. Peak labels imply peak numbers, with the subscripts indicating
the polarization directions.}

\end{figure}

Calculated EA spectra of the AFM and FM phases of QDs with $C_{2v}$
symmetry (GQD-40), and the ones with no symmetry (GQD-38 and GQD-48),
corresponding to transverse, longitudinal and diagonal electric field,
are presented in Figs. \ref{optic_asymm_unbal-transverse}, \ref{optic_asymm_unbal-longitudinal},
\ref{optic_asymm_unbal-diagonal}, \ref{optic_asymm_GQD-40-GQD-48-transverse},
\ref{optic_asymm_GQD-40-GQD-48-longitudinal} and \ref{optic_asymm_GQD-40-GQD-48-diagonal}.
Main features of the spectra are summarized below.

\begin{enumerate}
\item The EA spectra of both the AFM and FM phases of QDs of $C_{2v}$ (GQD-40),
and totally asymmetric QDs (GQD-38 and GQD-48), exhibit spin splitting
(Figs. \ref{optic_asymm_unbal-transverse} and \ref{optic_asymm_GQD-40-GQD-48-transverse}),
in the presence of transverse electric field. In case of the AFM state,
the EA spectra for $\alpha$ and $\beta$ spins are shifted in opposite
directions with the increasing field strength, with the spectra for
$\alpha/\beta$ spins exhibiting blue/red shifts, with respect to
the absorption spectra in the absence of electric field. However,
for the FM phase, the EA spectra for both types of spins get red shifted
with the increasing field. This marked difference exhibited by the
EA spectrum of AFM and FM states can be used to characterize the magnetic
ground states of QDs of lower symmetries. 
\item Absorption spectra of FM phase of completely asymmetric GQDs (GQD-38
and GQD-48) exhibit maximum sensitivity towards electric fields, and
split for two spin orientations when exposed to an in-plane electric
field in any direction (Figs. \ref{optic_asymm_unbal-transverse},
\ref{optic_asymm_unbal-longitudinal}, \ref{optic_asymm_unbal-diagonal},
\ref{optic_asymm_GQD-40-GQD-48-transverse}, \ref{optic_asymm_GQD-40-GQD-48-longitudinal}
and \ref{optic_asymm_GQD-40-GQD-48-diagonal}). However, the EA spectra
of the AFM and FM states of QDs of $C_{2v}$ symmetry (GQD-40), similar
to the behavior discussed earlier for QDs with $D_{2h}$ symmetry,
does not exhibit a spin-sensitive split on application of a longitudinal
electric field (Fig. \ref{optic_asymm_GQD-40-GQD-48-longitudinal}).
This distinctive feature displayed by the EA spectrum can be used
to distinguish the low symmetry QDs, from totally asymmetric ones. 
\item In addition, when the magnetic ground states of GQD-38 and GQD-40
undergo a phase transition to the NM phase, the spin-dependent optical
splitting vanishes. For the AFM phase, the field-induced splitting
of the EA spectrum for two spin orientations is more pronounced in
the lower energy region, because, the higher energy peaks are due
to excitations from orbitals further away from the Fermi level, which
exhibit decreasing spin polarity. For all the QDs studied, the peak
patterns for the two spin orientations for AFM/FM states are quite
distinct, along with their relative intensities, even at low values
of the applied electric field. Thus, it is possible to identify the
magnetic ground state, as well as the energy states attained after
electric-field driven phase transitions, by means of EA spectroscopy. 
\end{enumerate}
\begin{figure}[H]

\includegraphics[scale=0.4]{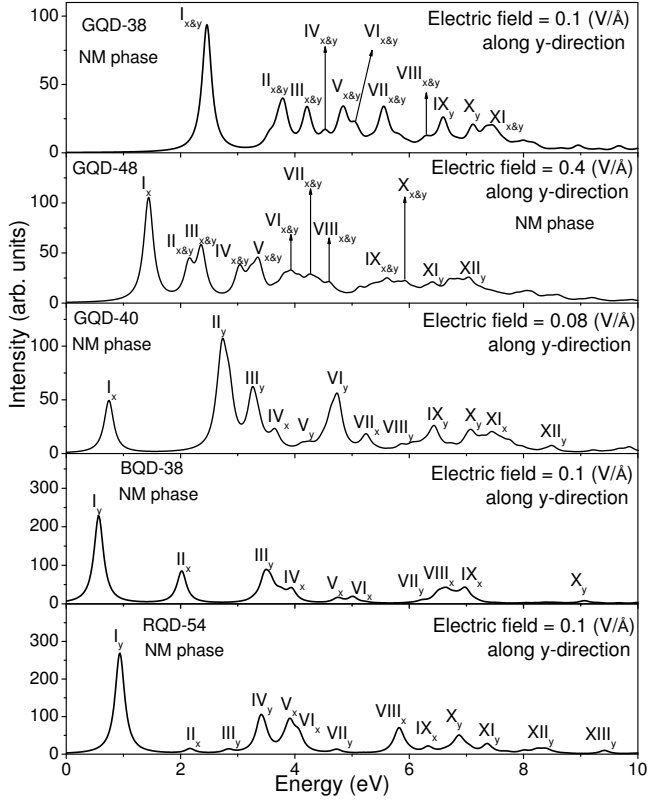}\caption{Computed optical absorption spectra for NM phases of GQD-38, GQD-48,
GQD-40, BQD-38, and RQD-54 in the presence of an in-plane transverse
electric field. The computed spectrum has been broadened with a uniform
line-width of 0.1 eV. \label{fig:Computed-optical-absorption-NM}Peak
labels imply peak numbers, with the subscripts indicating the polarization
directions.}

\end{figure}

The EA spectra for the NM phase, in the presence of an in-plane transverse
electric field, for all the QDs considered in this work has been given
in Fig. \ref{fig:Computed-optical-absorption-NM}. It is observed
that the EA spectra for the NM phase is drastically different from
the EA spectra of the magnetic phases (AFM and FM). Thus, we expect
that EA spectra can be used as to successfully differentiate various
magnetic and non-magnetic states of graphene quantum dots. 

\section{Conc\label{sec:Conclusions}lusions}

In this work, calculations on different magnetic states (AFM and FM)
of graphene QDs have indicated that energy gaps of these magnetic
states exhibit spin dependence. Also, the manner in which the spin-dependent
energy gaps vary, when exposed to a suitably aligned electric field,
is distinct for the AFM and FM configurations of QDs. This empowers
us to manipulate the band gaps by an external electric field, which
is of fundamental significance in spintronics. Additionally, our studies
have elucidated the correlation between spin density, and optical
band-gap splitting. It has been shown that energy band-gap splitting
for the AFM phase arises due to spatial localization of spin density
corresponding to the two distinct spin orientations on the opposite
edges of the quantum dots. In contrast, higher concentration of spin
density associated with same spin in one region of the QD, as compared
to the other regions, is responsible for band-gap splitting in the
FM state. We have then shown that the magnetic states of QDs undergo
phase transitions on application of an electric field. Furthermore,
we have demonstrated that the EA spectra of different magnetic states
of QDs have distinct footprints (peak pattern as well as nature of
peak shifts). Thus, by studying the variation of the EA spectra as
a function of the field strength, one can efficiently distinguish
different magnetic states of QDs. We hope that our findings will create
ways to realize spintronic devices based on graphene QDs, in the near
future. 

\bibliographystyle{apsrev4-1}
\bibliography{Graphenemag}

\begin{thebibliography}{43}%
\makeatletter
\providecommand \@ifxundefined [1]{%
 \@ifx{#1\undefined}
}%
\providecommand \@ifnum [1]{%
 \ifnum #1\expandafter \@firstoftwo
 \else \expandafter \@secondoftwo
 \fi
}%
\providecommand \@ifx [1]{%
 \ifx #1\expandafter \@firstoftwo
 \else \expandafter \@secondoftwo
 \fi
}%
\providecommand \natexlab [1]{#1}%
\providecommand \enquote  [1]{``#1''}%
\providecommand \bibnamefont  [1]{#1}%
\providecommand \bibfnamefont [1]{#1}%
\providecommand \citenamefont [1]{#1}%
\providecommand \href@noop [0]{\@secondoftwo}%
\providecommand \href [0]{\begingroup \@sanitize@url \@href}%
\providecommand \@href[1]{\@@startlink{#1}\@@href}%
\providecommand \@@href[1]{\endgroup#1\@@endlink}%
\providecommand \@sanitize@url [0]{\catcode `\\12\catcode `\$12\catcode
  `\&12\catcode `\#12\catcode `\^12\catcode `\_12\catcode `\%12\relax}%
\providecommand \@@startlink[1]{}%
\providecommand \@@endlink[0]{}%
\providecommand \url  [0]{\begingroup\@sanitize@url \@url }%
\providecommand \@url [1]{\endgroup\@href {#1}{\urlprefix }}%
\providecommand \urlprefix  [0]{URL }%
\providecommand \Eprint [0]{\href }%
\providecommand \doibase [0]{http://dx.doi.org/}%
\providecommand \selectlanguage [0]{\@gobble}%
\providecommand \bibinfo  [0]{\@secondoftwo}%
\providecommand \bibfield  [0]{\@secondoftwo}%
\providecommand \translation [1]{[#1]}%
\providecommand \BibitemOpen [0]{}%
\providecommand \bibitemStop [0]{}%
\providecommand \bibitemNoStop [0]{.\EOS\space}%
\providecommand \EOS [0]{\spacefactor3000\relax}%
\providecommand \BibitemShut  [1]{\csname bibitem#1\endcsname}%
\let\auto@bib@innerbib\@empty
%</preamble>
\bibitem [{\citenamefont {Son}\ \emph {et~al.}(2006)\citenamefont {Son},
  \citenamefont {Cohen},\ and\ \citenamefont {Louie}}]{Son}%
  \BibitemOpen
  \bibfield  {author} {\bibinfo {author} {\bibfnamefont {Y.-W.}\ \bibnamefont
  {Son}}, \bibinfo {author} {\bibfnamefont {M.~L.}\ \bibnamefont {Cohen}}, \
  and\ \bibinfo {author} {\bibfnamefont {S.~G.}\ \bibnamefont {Louie}},\ }\href
  {http://dx.doi.org/10.1038/nature05180} {\bibfield  {journal} {\bibinfo
  {journal} {Nature}\ }\textbf {\bibinfo {volume} {444}},\ \bibinfo {pages}
  {347} (\bibinfo {year} {2006})}\BibitemShut {NoStop}%
\bibitem [{\citenamefont {Agapito}\ \emph {et~al.}(2010)\citenamefont
  {Agapito}, \citenamefont {Kioussis},\ and\ \citenamefont
  {Kaxiras}}]{PhysRevB.82.201411Agapito}%
  \BibitemOpen
  \bibfield  {author} {\bibinfo {author} {\bibfnamefont {L.~A.}\ \bibnamefont
  {Agapito}}, \bibinfo {author} {\bibfnamefont {N.}~\bibnamefont {Kioussis}}, \
  and\ \bibinfo {author} {\bibfnamefont {E.}~\bibnamefont {Kaxiras}},\ }\href
  {\doibase 10.1103/PhysRevB.82.201411} {\bibfield  {journal} {\bibinfo
  {journal} {Phys. Rev. B}\ }\textbf {\bibinfo {volume} {82}},\ \bibinfo
  {pages} {201411} (\bibinfo {year} {2010})}\BibitemShut {NoStop}%
\bibitem [{\citenamefont {Zheng}\ and\ \citenamefont
  {Duley}(2008{\natexlab{a}})}]{ZhengdopedPhysRevB.78.155118}%
  \BibitemOpen
  \bibfield  {author} {\bibinfo {author} {\bibfnamefont {H.}~\bibnamefont
  {Zheng}}\ and\ \bibinfo {author} {\bibfnamefont {W.}~\bibnamefont {Duley}},\
  }\href {\doibase 10.1103/PhysRevB.78.155118} {\bibfield  {journal} {\bibinfo
  {journal} {Phys. Rev. B}\ }\textbf {\bibinfo {volume} {78}},\ \bibinfo
  {pages} {155118} (\bibinfo {year} {2008}{\natexlab{a}})}\BibitemShut
  {NoStop}%
\bibitem [{\citenamefont {Bhowmick}\ and\ \citenamefont
  {Shenoy}(2008)}]{Bhowmick:/content/aip/journal/jcp/128/24/10.1063/1.2943678}%
  \BibitemOpen
  \bibfield  {author} {\bibinfo {author} {\bibfnamefont {S.}~\bibnamefont
  {Bhowmick}}\ and\ \bibinfo {author} {\bibfnamefont {V.~B.}\ \bibnamefont
  {Shenoy}},\ }\href
  {http://scitation.aip.org/content/aip/journal/jcp/128/24/10.1063/1.2943678}
  {\bibfield  {journal} {\bibinfo  {journal} {The Journal of Chemical Physics}\
  }\textbf {\bibinfo {volume} {128}},\ \bibinfo {eid} {244717} (\bibinfo {year}
  {2008})}\BibitemShut {NoStop}%
\bibitem [{\citenamefont {Zhou}\ \emph {et~al.}(2013)\citenamefont {Zhou},
  \citenamefont {Sheng},\ and\ \citenamefont
  {Xu}}]{:/content/aip/journal/apl/103/13/10.1063/1.4821954Zhou}%
  \BibitemOpen
  \bibfield  {author} {\bibinfo {author} {\bibfnamefont {A.}~\bibnamefont
  {Zhou}}, \bibinfo {author} {\bibfnamefont {W.}~\bibnamefont {Sheng}}, \ and\
  \bibinfo {author} {\bibfnamefont {S.~J.}\ \bibnamefont {Xu}},\ }\href
  {http://scitation.aip.org/content/aip/journal/apl/103/13/10.1063/1.4821954}
  {\bibfield  {journal} {\bibinfo  {journal} {Applied Physics Letters}\
  }\textbf {\bibinfo {volume} {103}},\ \bibinfo {eid} {133103} (\bibinfo {year}
  {2013})}\BibitemShut {NoStop}%
\bibitem [{\citenamefont {Ma}\ and\ \citenamefont
  {Li}(2012)}]{PhysRevB.86.045449Ma}%
  \BibitemOpen
  \bibfield  {author} {\bibinfo {author} {\bibfnamefont {W.-L.}\ \bibnamefont
  {Ma}}\ and\ \bibinfo {author} {\bibfnamefont {S.-S.}\ \bibnamefont {Li}},\
  }\href {\doibase 10.1103/PhysRevB.86.045449} {\bibfield  {journal} {\bibinfo
  {journal} {Phys. Rev. B}\ }\textbf {\bibinfo {volume} {86}},\ \bibinfo
  {pages} {045449} (\bibinfo {year} {2012})}\BibitemShut {NoStop}%
\bibitem [{\citenamefont {Sahin}\ \emph {et~al.}(2010)\citenamefont {Sahin},
  \citenamefont {Senger},\ and\ \citenamefont
  {Ciraci}}]{:/content/aip/journal/jap/108/7/10.1063/1.3489919Sahin}%
  \BibitemOpen
  \bibfield  {author} {\bibinfo {author} {\bibfnamefont {H.}~\bibnamefont
  {Sahin}}, \bibinfo {author} {\bibfnamefont {R.~T.}\ \bibnamefont {Senger}}, \
  and\ \bibinfo {author} {\bibfnamefont {S.}~\bibnamefont {Ciraci}},\ }\href
  {http://scitation.aip.org/content/aip/journal/jap/108/7/10.1063/1.3489919}
  {\bibfield  {journal} {\bibinfo  {journal} {Journal of Applied Physics}\
  }\textbf {\bibinfo {volume} {108}},\ \bibinfo {eid} {074301} (\bibinfo {year}
  {2010})}\BibitemShut {NoStop}%
\bibitem [{\citenamefont {Wang}\ \emph {et~al.}(2008)\citenamefont {Wang},
  \citenamefont {Meng},\ and\ \citenamefont
  {Kaxiras}}]{doi:10.1021/nl072548aWang}%
  \BibitemOpen
  \bibfield  {author} {\bibinfo {author} {\bibfnamefont {W.~L.}\ \bibnamefont
  {Wang}}, \bibinfo {author} {\bibfnamefont {S.}~\bibnamefont {Meng}}, \ and\
  \bibinfo {author} {\bibfnamefont {E.}~\bibnamefont {Kaxiras}},\ }\href
  {http://dx.doi.org/10.1021/nl072548a} {\bibfield  {journal} {\bibinfo
  {journal} {Nano Letters}\ }\textbf {\bibinfo {volume} {8}},\ \bibinfo {pages}
  {241} (\bibinfo {year} {2008})}\BibitemShut {NoStop}%
\bibitem [{\citenamefont {Feldner}\ \emph {et~al.}(2010)\citenamefont
  {Feldner}, \citenamefont {Meng}, \citenamefont {Honecker}, \citenamefont
  {Cabra}, \citenamefont {Wessel},\ and\ \citenamefont
  {Assaad}}]{FeldnerPhysRevB.81.115416}%
  \BibitemOpen
  \bibfield  {author} {\bibinfo {author} {\bibfnamefont {H.}~\bibnamefont
  {Feldner}}, \bibinfo {author} {\bibfnamefont {Z.~Y.}\ \bibnamefont {Meng}},
  \bibinfo {author} {\bibfnamefont {A.}~\bibnamefont {Honecker}}, \bibinfo
  {author} {\bibfnamefont {D.}~\bibnamefont {Cabra}}, \bibinfo {author}
  {\bibfnamefont {S.}~\bibnamefont {Wessel}}, \ and\ \bibinfo {author}
  {\bibfnamefont {F.~F.}\ \bibnamefont {Assaad}},\ }\href
  {http://link.aps.org/doi/10.1103/PhysRevB.81.115416} {\bibfield  {journal}
  {\bibinfo  {journal} {Phys. Rev. B}\ }\textbf {\bibinfo {volume} {81}},\
  \bibinfo {pages} {115416} (\bibinfo {year} {2010})}\BibitemShut {NoStop}%
\bibitem [{\citenamefont {Zhang}\ \emph {et~al.}(2008)\citenamefont {Zhang},
  \citenamefont {Chang},\ and\ \citenamefont
  {Peeters}}]{PhysRevB.77.235411Zhang}%
  \BibitemOpen
  \bibfield  {author} {\bibinfo {author} {\bibfnamefont {Z.~Z.}\ \bibnamefont
  {Zhang}}, \bibinfo {author} {\bibfnamefont {K.}~\bibnamefont {Chang}}, \ and\
  \bibinfo {author} {\bibfnamefont {F.~M.}\ \bibnamefont {Peeters}},\ }\href
  {http://link.aps.org/doi/10.1103/PhysRevB.77.235411} {\bibfield  {journal}
  {\bibinfo  {journal} {Phys. Rev. B}\ }\textbf {\bibinfo {volume} {77}},\
  \bibinfo {pages} {235411} (\bibinfo {year} {2008})}\BibitemShut {NoStop}%
\bibitem [{\citenamefont {Zheng}\ and\ \citenamefont
  {Duley}(2008{\natexlab{b}})}]{PhysRevB.78.045421Zheng}%
  \BibitemOpen
  \bibfield  {author} {\bibinfo {author} {\bibfnamefont {H.}~\bibnamefont
  {Zheng}}\ and\ \bibinfo {author} {\bibfnamefont {W.}~\bibnamefont {Duley}},\
  }\href {http://link.aps.org/doi/10.1103/PhysRevB.78.045421} {\bibfield
  {journal} {\bibinfo  {journal} {Phys. Rev. B}\ }\textbf {\bibinfo {volume}
  {78}},\ \bibinfo {pages} {045421} (\bibinfo {year}
  {2008}{\natexlab{b}})}\BibitemShut {NoStop}%
\bibitem [{\citenamefont {Zarenia}\ \emph {et~al.}(2011)\citenamefont
  {Zarenia}, \citenamefont {Chaves}, \citenamefont {Farias},\ and\
  \citenamefont {Peeters}}]{PhysRevB.84.245403Zarenia}%
  \BibitemOpen
  \bibfield  {author} {\bibinfo {author} {\bibfnamefont {M.}~\bibnamefont
  {Zarenia}}, \bibinfo {author} {\bibfnamefont {A.}~\bibnamefont {Chaves}},
  \bibinfo {author} {\bibfnamefont {G.~A.}\ \bibnamefont {Farias}}, \ and\
  \bibinfo {author} {\bibfnamefont {F.~M.}\ \bibnamefont {Peeters}},\ }\href
  {http://link.aps.org/doi/10.1103/PhysRevB.84.245403} {\bibfield  {journal}
  {\bibinfo  {journal} {Phys. Rev. B}\ }\textbf {\bibinfo {volume} {84}},\
  \bibinfo {pages} {245403} (\bibinfo {year} {2011})}\BibitemShut {NoStop}%
\bibitem [{\citenamefont {G\"u\ifmmode~\mbox{\c{c}}\else \c{c}\fi{}l\"u}\ and\
  \citenamefont {Hawrylak}(2013)}]{PhysRevB.87.035425Guclu}%
  \BibitemOpen
  \bibfield  {author} {\bibinfo {author} {\bibfnamefont {A.~D.}\ \bibnamefont
  {G\"u\ifmmode~\mbox{\c{c}}\else \c{c}\fi{}l\"u}}\ and\ \bibinfo {author}
  {\bibfnamefont {P.}~\bibnamefont {Hawrylak}},\ }\href
  {http://link.aps.org/doi/10.1103/PhysRevB.87.035425} {\bibfield  {journal}
  {\bibinfo  {journal} {Phys. Rev. B}\ }\textbf {\bibinfo {volume} {87}},\
  \bibinfo {pages} {035425} (\bibinfo {year} {2013})}\BibitemShut {NoStop}%
\bibitem [{\citenamefont {Fern\'andez-Rossier}\ and\ \citenamefont
  {Palacios}(2007)}]{PhysRevLett.99.177204Fernandes}%
  \BibitemOpen
  \bibfield  {author} {\bibinfo {author} {\bibfnamefont {J.}~\bibnamefont
  {Fern\'andez-Rossier}}\ and\ \bibinfo {author} {\bibfnamefont {J.~J.}\
  \bibnamefont {Palacios}},\ }\href
  {http://link.aps.org/doi/10.1103/PhysRevLett.99.177204} {\bibfield  {journal}
  {\bibinfo  {journal} {Phys. Rev. Lett.}\ }\textbf {\bibinfo {volume} {99}},\
  \bibinfo {pages} {177204} (\bibinfo {year} {2007})}\BibitemShut {NoStop}%
\bibitem [{\citenamefont {Agapito}\ and\ \citenamefont
  {Kioussis}(2011)}]{Agapitodoi:10.1021/jp1096234}%
  \BibitemOpen
  \bibfield  {author} {\bibinfo {author} {\bibfnamefont {L.~A.}\ \bibnamefont
  {Agapito}}\ and\ \bibinfo {author} {\bibfnamefont {N.}~\bibnamefont
  {Kioussis}},\ }\href {http://dx.doi.org/10.1021/jp1096234} {\bibfield
  {journal} {\bibinfo  {journal} {The Journal of Physical Chemistry C}\
  }\textbf {\bibinfo {volume} {115}},\ \bibinfo {pages} {2874} (\bibinfo {year}
  {2011})}\BibitemShut {NoStop}%
\bibitem [{\citenamefont {G\"u\ifmmode~\mbox{\c{c}}\else \c{c}\fi{}l\"u}\ \emph
  {et~al.}(2010)\citenamefont {G\"u\ifmmode~\mbox{\c{c}}\else \c{c}\fi{}l\"u},
  \citenamefont {Potasz},\ and\ \citenamefont {Hawrylak}}]{efieldmag-hawrylak}%
  \BibitemOpen
  \bibfield  {author} {\bibinfo {author} {\bibfnamefont {A.~D.}\ \bibnamefont
  {G\"u\ifmmode~\mbox{\c{c}}\else \c{c}\fi{}l\"u}}, \bibinfo {author}
  {\bibfnamefont {P.}~\bibnamefont {Potasz}}, \ and\ \bibinfo {author}
  {\bibfnamefont {P.}~\bibnamefont {Hawrylak}},\ }\href {\doibase
  10.1103/PhysRevB.82.155445} {\bibfield  {journal} {\bibinfo  {journal} {Phys.
  Rev. B}\ }\textbf {\bibinfo {volume} {82}},\ \bibinfo {pages} {155445}
  (\bibinfo {year} {2010})}\BibitemShut {NoStop}%
\bibitem [{\citenamefont {Sza\l{}owski}(2014)}]{efieldmag-karol}%
  \BibitemOpen
  \bibfield  {author} {\bibinfo {author} {\bibfnamefont {K.}~\bibnamefont
  {Sza\l{}owski}},\ }\href {\doibase 10.1103/PhysRevB.90.085410} {\bibfield
  {journal} {\bibinfo  {journal} {Phys. Rev. B}\ }\textbf {\bibinfo {volume}
  {90}},\ \bibinfo {pages} {085410} (\bibinfo {year} {2014})}\BibitemShut
  {NoStop}%
\bibitem [{\citenamefont {Tombros}\ \emph {et~al.}(2007)\citenamefont
  {Tombros}, \citenamefont {Jozsa}, \citenamefont {Popinciuc}, \citenamefont
  {Jonkman},\ and\ \citenamefont {van Wees}}]{Tombros}%
  \BibitemOpen
  \bibfield  {author} {\bibinfo {author} {\bibfnamefont {N.}~\bibnamefont
  {Tombros}}, \bibinfo {author} {\bibfnamefont {C.}~\bibnamefont {Jozsa}},
  \bibinfo {author} {\bibfnamefont {M.}~\bibnamefont {Popinciuc}}, \bibinfo
  {author} {\bibfnamefont {H.~T.}\ \bibnamefont {Jonkman}}, \ and\ \bibinfo
  {author} {\bibfnamefont {B.~J.}\ \bibnamefont {van Wees}},\ }\href
  {http://dx.doi.org/10.1038/nature06037} {\bibfield  {journal} {\bibinfo
  {journal} {Nature}\ }\textbf {\bibinfo {volume} {448}},\ \bibinfo {pages}
  {571} (\bibinfo {year} {2007})}\BibitemShut {NoStop}%
\bibitem [{\citenamefont {Yazyev}\ and\ \citenamefont
  {Katsnelson}(2008)}]{PhysRevLett.100.047209Yazyev}%
  \BibitemOpen
  \bibfield  {author} {\bibinfo {author} {\bibfnamefont {O.~V.}\ \bibnamefont
  {Yazyev}}\ and\ \bibinfo {author} {\bibfnamefont {M.~I.}\ \bibnamefont
  {Katsnelson}},\ }\href
  {http://link.aps.org/doi/10.1103/PhysRevLett.100.047209} {\bibfield
  {journal} {\bibinfo  {journal} {Phys. Rev. Lett.}\ }\textbf {\bibinfo
  {volume} {100}},\ \bibinfo {pages} {047209} (\bibinfo {year}
  {2008})}\BibitemShut {NoStop}%
\bibitem [{\citenamefont {Han}\ \emph {et~al.}(2009)\citenamefont {Han},
  \citenamefont {Pi}, \citenamefont {Bao}, \citenamefont {McCreary},
  \citenamefont {Li}, \citenamefont {Wang}, \citenamefont {Lau},\ and\
  \citenamefont
  {Kawakami}}]{:/content/aip/journal/apl/94/22/10.1063/1.3147203Han}%
  \BibitemOpen
  \bibfield  {author} {\bibinfo {author} {\bibfnamefont {W.}~\bibnamefont
  {Han}}, \bibinfo {author} {\bibfnamefont {K.}~\bibnamefont {Pi}}, \bibinfo
  {author} {\bibfnamefont {W.}~\bibnamefont {Bao}}, \bibinfo {author}
  {\bibfnamefont {K.~M.}\ \bibnamefont {McCreary}}, \bibinfo {author}
  {\bibfnamefont {Y.}~\bibnamefont {Li}}, \bibinfo {author} {\bibfnamefont
  {W.~H.}\ \bibnamefont {Wang}}, \bibinfo {author} {\bibfnamefont {C.~N.}\
  \bibnamefont {Lau}}, \ and\ \bibinfo {author} {\bibfnamefont {R.~K.}\
  \bibnamefont {Kawakami}},\ }\href
  {http://scitation.aip.org/content/aip/journal/apl/94/22/10.1063/1.3147203}
  {\bibfield  {journal} {\bibinfo  {journal} {Applied Physics Letters}\
  }\textbf {\bibinfo {volume} {94}},\ \bibinfo {eid} {222109} (\bibinfo {year}
  {2009})}\BibitemShut {NoStop}%
\bibitem [{\citenamefont {Yang}\ \emph {et~al.}(2011)\citenamefont {Yang},
  \citenamefont {Balakrishnan}, \citenamefont {Volmer}, \citenamefont {Avsar},
  \citenamefont {Jaiswal}, \citenamefont {Samm}, \citenamefont {Ali},
  \citenamefont {Pachoud}, \citenamefont {Zeng}, \citenamefont {Popinciuc},
  \citenamefont {G\"untherodt}, \citenamefont {Beschoten},\ and\ \citenamefont
  {\"Ozyilmaz}}]{PhysRevLett.107.047206Yang}%
  \BibitemOpen
  \bibfield  {author} {\bibinfo {author} {\bibfnamefont {T.-Y.}\ \bibnamefont
  {Yang}}, \bibinfo {author} {\bibfnamefont {J.}~\bibnamefont {Balakrishnan}},
  \bibinfo {author} {\bibfnamefont {F.}~\bibnamefont {Volmer}}, \bibinfo
  {author} {\bibfnamefont {A.}~\bibnamefont {Avsar}}, \bibinfo {author}
  {\bibfnamefont {M.}~\bibnamefont {Jaiswal}}, \bibinfo {author} {\bibfnamefont
  {J.}~\bibnamefont {Samm}}, \bibinfo {author} {\bibfnamefont {S.~R.}\
  \bibnamefont {Ali}}, \bibinfo {author} {\bibfnamefont {A.}~\bibnamefont
  {Pachoud}}, \bibinfo {author} {\bibfnamefont {M.}~\bibnamefont {Zeng}},
  \bibinfo {author} {\bibfnamefont {M.}~\bibnamefont {Popinciuc}}, \bibinfo
  {author} {\bibfnamefont {G.}~\bibnamefont {G\"untherodt}}, \bibinfo {author}
  {\bibfnamefont {B.}~\bibnamefont {Beschoten}}, \ and\ \bibinfo {author}
  {\bibfnamefont {B.}~\bibnamefont {\"Ozyilmaz}},\ }\href
  {http://link.aps.org/doi/10.1103/PhysRevLett.107.047206} {\bibfield
  {journal} {\bibinfo  {journal} {Phys. Rev. Lett.}\ }\textbf {\bibinfo
  {volume} {107}},\ \bibinfo {pages} {047206} (\bibinfo {year}
  {2011})}\BibitemShut {NoStop}%
\bibitem [{\citenamefont {Chen}\ \emph {et~al.}(2013)\citenamefont {Chen},
  \citenamefont {Guo}, \citenamefont {Li}, \citenamefont {Zhang}, \citenamefont
  {Lin}, \citenamefont {Huang}, \citenamefont {Jin},\ and\ \citenamefont
  {Chen}}]{Chen_nature}%
  \BibitemOpen
  \bibfield  {author} {\bibinfo {author} {\bibfnamefont {L.}~\bibnamefont
  {Chen}}, \bibinfo {author} {\bibfnamefont {L.}~\bibnamefont {Guo}}, \bibinfo
  {author} {\bibfnamefont {Z.}~\bibnamefont {Li}}, \bibinfo {author}
  {\bibfnamefont {H.}~\bibnamefont {Zhang}}, \bibinfo {author} {\bibfnamefont
  {J.}~\bibnamefont {Lin}}, \bibinfo {author} {\bibfnamefont {J.}~\bibnamefont
  {Huang}}, \bibinfo {author} {\bibfnamefont {S.}~\bibnamefont {Jin}}, \ and\
  \bibinfo {author} {\bibfnamefont {X.}~\bibnamefont {Chen}},\ }\href
  {http://dx.doi.org/10.1038/srep02599} {\bibfield  {journal} {\bibinfo
  {journal} {Sci. Rep.}\ }\textbf {\bibinfo {volume} {3}},\ \bibinfo {pages}
  {2599} (\bibinfo {year} {2013})}\BibitemShut {NoStop}%
\bibitem [{\citenamefont {Magda}\ \emph {et~al.}(2014)\citenamefont {Magda},
  \citenamefont {Jin}, \citenamefont {Hagymasi}, \citenamefont {Vancso},
  \citenamefont {Osvath}, \citenamefont {Nemes-Incze}, \citenamefont {Hwang},
  \citenamefont {Biro},\ and\ \citenamefont {Tapaszto}}]{Magda}%
  \BibitemOpen
  \bibfield  {author} {\bibinfo {author} {\bibfnamefont {G.~Z.}\ \bibnamefont
  {Magda}}, \bibinfo {author} {\bibfnamefont {X.}~\bibnamefont {Jin}}, \bibinfo
  {author} {\bibfnamefont {I.}~\bibnamefont {Hagymasi}}, \bibinfo {author}
  {\bibfnamefont {P.}~\bibnamefont {Vancso}}, \bibinfo {author} {\bibfnamefont
  {Z.}~\bibnamefont {Osvath}}, \bibinfo {author} {\bibfnamefont
  {P.}~\bibnamefont {Nemes-Incze}}, \bibinfo {author} {\bibfnamefont
  {C.}~\bibnamefont {Hwang}}, \bibinfo {author} {\bibfnamefont {L.~P.}\
  \bibnamefont {Biro}}, \ and\ \bibinfo {author} {\bibfnamefont
  {L.}~\bibnamefont {Tapaszto}},\ }\href
  {http://dx.doi.org/10.1038/nature13831} {\bibfield  {journal} {\bibinfo
  {journal} {Nature}\ }\textbf {\bibinfo {volume} {514}},\ \bibinfo {pages}
  {608} (\bibinfo {year} {2014})}\BibitemShut {NoStop}%
\bibitem [{\citenamefont {Kiguchi}\ \emph {et~al.}(2011)\citenamefont
  {Kiguchi}, \citenamefont {Takai}, \citenamefont {Joly}, \citenamefont
  {Enoki}, \citenamefont {Sumii},\ and\ \citenamefont
  {Amemiya}}]{PhysRevB.84.045421_Kiguchi}%
  \BibitemOpen
  \bibfield  {author} {\bibinfo {author} {\bibfnamefont {M.}~\bibnamefont
  {Kiguchi}}, \bibinfo {author} {\bibfnamefont {K.}~\bibnamefont {Takai}},
  \bibinfo {author} {\bibfnamefont {V.~L.~J.}\ \bibnamefont {Joly}}, \bibinfo
  {author} {\bibfnamefont {T.}~\bibnamefont {Enoki}}, \bibinfo {author}
  {\bibfnamefont {R.}~\bibnamefont {Sumii}}, \ and\ \bibinfo {author}
  {\bibfnamefont {K.}~\bibnamefont {Amemiya}},\ }\href {\doibase
  10.1103/PhysRevB.84.045421} {\bibfield  {journal} {\bibinfo  {journal} {Phys.
  Rev. B}\ }\textbf {\bibinfo {volume} {84}},\ \bibinfo {pages} {045421}
  (\bibinfo {year} {2011})}\BibitemShut {NoStop}%
\bibitem [{\citenamefont {Suenaga}\ and\ \citenamefont
  {Koshino}(2010)}]{Suenaga_nature}%
  \BibitemOpen
  \bibfield  {author} {\bibinfo {author} {\bibfnamefont {K.}~\bibnamefont
  {Suenaga}}\ and\ \bibinfo {author} {\bibfnamefont {M.}~\bibnamefont
  {Koshino}},\ }\href {http://dx.doi.org/10.1038/nature09664} {\bibfield
  {journal} {\bibinfo  {journal} {Nature}\ }\textbf {\bibinfo {volume} {468}},\
  \bibinfo {pages} {1088} (\bibinfo {year} {2010})}\BibitemShut {NoStop}%
\bibitem [{\citenamefont {Gundra}\ and\ \citenamefont
  {Shukla}(2011{\natexlab{a}})}]{PhysRevB.83.075413_Gundra}%
  \BibitemOpen
  \bibfield  {author} {\bibinfo {author} {\bibfnamefont {K.}~\bibnamefont
  {Gundra}}\ and\ \bibinfo {author} {\bibfnamefont {A.}~\bibnamefont
  {Shukla}},\ }\href {\doibase 10.1103/PhysRevB.83.075413} {\bibfield
  {journal} {\bibinfo  {journal} {Phys. Rev. B}\ }\textbf {\bibinfo {volume}
  {83}},\ \bibinfo {pages} {075413} (\bibinfo {year}
  {2011}{\natexlab{a}})}\BibitemShut {NoStop}%
\bibitem [{\citenamefont {Pople}(1953)}]{ppp-pople}%
  \BibitemOpen
  \bibfield  {author} {\bibinfo {author} {\bibfnamefont {J.~A.}\ \bibnamefont
  {Pople}},\ }\href {\doibase 10.1039/TF9534901375} {\bibfield  {journal}
  {\bibinfo  {journal} {Trans. Faraday Soc.}\ }\textbf {\bibinfo {volume}
  {49}},\ \bibinfo {pages} {1375} (\bibinfo {year} {1953})}\BibitemShut
  {NoStop}%
\bibitem [{\citenamefont {Pariser}\ and\ \citenamefont
  {Parr}(1953)}]{ppp-pariser-parr}%
  \BibitemOpen
  \bibfield  {author} {\bibinfo {author} {\bibfnamefont {R.}~\bibnamefont
  {Pariser}}\ and\ \bibinfo {author} {\bibfnamefont {R.~G.}\ \bibnamefont
  {Parr}},\ }\href {\doibase http://dx.doi.org/10.1063/1.1699030} {\bibfield
  {journal} {\bibinfo  {journal} {J. Chem. Phys.}\ }\textbf {\bibinfo {volume}
  {21}},\ \bibinfo {pages} {767} (\bibinfo {year} {1953})}\BibitemShut
  {NoStop}%
\bibitem [{\citenamefont {Sony}\ and\ \citenamefont
  {Shukla}(2005)}]{PhysRevB.71.165204Priya_t0}%
  \BibitemOpen
  \bibfield  {author} {\bibinfo {author} {\bibfnamefont {P.}~\bibnamefont
  {Sony}}\ and\ \bibinfo {author} {\bibfnamefont {A.}~\bibnamefont {Shukla}},\
  }\href {\doibase 10.1103/PhysRevB.71.165204} {\bibfield  {journal} {\bibinfo
  {journal} {Phys. Rev. B}\ }\textbf {\bibinfo {volume} {71}},\ \bibinfo
  {pages} {165204} (\bibinfo {year} {2005})}\BibitemShut {NoStop}%
\bibitem [{\citenamefont {Aryanpour}\ \emph
  {et~al.}(2014{\natexlab{a}})\citenamefont {Aryanpour}, \citenamefont
  {Shukla},\ and\ \citenamefont
  {Mazumdar}}]{:/content/aip/journal/jcp/140/10/10.1063/1.4867363Aryanpour}%
  \BibitemOpen
  \bibfield  {author} {\bibinfo {author} {\bibfnamefont {K.}~\bibnamefont
  {Aryanpour}}, \bibinfo {author} {\bibfnamefont {A.}~\bibnamefont {Shukla}}, \
  and\ \bibinfo {author} {\bibfnamefont {S.}~\bibnamefont {Mazumdar}},\ }\href
  {\doibase http://dx.doi.org/10.1063/1.4867363} {\bibfield  {journal}
  {\bibinfo  {journal} {The Journal of Chemical Physics}\ }\textbf {\bibinfo
  {volume} {140}},\ \bibinfo {eid} {104301} (\bibinfo {year}
  {2014}{\natexlab{a}})}\BibitemShut {NoStop}%
\bibitem [{\citenamefont {Basak}\ \emph {et~al.}(2015)\citenamefont {Basak},
  \citenamefont {Chakraborty},\ and\ \citenamefont {Shukla}}]{Tista}%
  \BibitemOpen
  \bibfield  {author} {\bibinfo {author} {\bibfnamefont {T.}~\bibnamefont
  {Basak}}, \bibinfo {author} {\bibfnamefont {H.}~\bibnamefont {Chakraborty}},
  \ and\ \bibinfo {author} {\bibfnamefont {A.}~\bibnamefont {Shukla}},\ }\href
  {\doibase 10.1103/PhysRevB.92.205404} {\bibfield  {journal} {\bibinfo
  {journal} {Phys. Rev. B}\ }\textbf {\bibinfo {volume} {92}},\ \bibinfo
  {pages} {205404} (\bibinfo {year} {2015})}\BibitemShut {NoStop}%
\bibitem [{\citenamefont {Ohno}(1964)}]{Theor.chim.act.2Ohno}%
  \BibitemOpen
  \bibfield  {author} {\bibinfo {author} {\bibfnamefont {K.}~\bibnamefont
  {Ohno}},\ }\href {\doibase 10.1007/BF00528281} {\bibfield  {journal}
  {\bibinfo  {journal} {Theoretica chimica acta}\ }\textbf {\bibinfo {volume}
  {2}},\ \bibinfo {pages} {219} (\bibinfo {year} {1964})}\BibitemShut {NoStop}%
\bibitem [{\citenamefont {Chandross}\ and\ \citenamefont
  {Mazumdar}(1997)}]{PhysRevB.55.1497Chandross}%
  \BibitemOpen
  \bibfield  {author} {\bibinfo {author} {\bibfnamefont {M.}~\bibnamefont
  {Chandross}}\ and\ \bibinfo {author} {\bibfnamefont {S.}~\bibnamefont
  {Mazumdar}},\ }\href {\doibase 10.1103/PhysRevB.55.1497} {\bibfield
  {journal} {\bibinfo  {journal} {Phys. Rev. B}\ }\textbf {\bibinfo {volume}
  {55}},\ \bibinfo {pages} {1497} (\bibinfo {year} {1997})}\BibitemShut
  {NoStop}%
\bibitem [{\citenamefont {Sony}\ and\ \citenamefont
  {Shukla}(2009)}]{:/content/aip/journal/jcp/131/1/10.1063/1.3159670Priyaanthracene}%
  \BibitemOpen
  \bibfield  {author} {\bibinfo {author} {\bibfnamefont {P.}~\bibnamefont
  {Sony}}\ and\ \bibinfo {author} {\bibfnamefont {A.}~\bibnamefont {Shukla}},\
  }\href {\doibase http://dx.doi.org/10.1063/1.3159670} {\bibfield  {journal}
  {\bibinfo  {journal} {The Journal of Chemical Physics}\ }\textbf {\bibinfo
  {volume} {131}},\ \bibinfo {eid} {014302} (\bibinfo {year}
  {2009})}\BibitemShut {NoStop}%
\bibitem [{\citenamefont {Chakraborty}\ and\ \citenamefont
  {Shukla}(2013)}]{doi10.1021/jp408535u}%
  \BibitemOpen
  \bibfield  {author} {\bibinfo {author} {\bibfnamefont {H.}~\bibnamefont
  {Chakraborty}}\ and\ \bibinfo {author} {\bibfnamefont {A.}~\bibnamefont
  {Shukla}},\ }\href {\doibase 10.1021/jp408535u} {\bibfield  {journal}
  {\bibinfo  {journal} {The Journal of Physical Chemistry A}\ }\textbf
  {\bibinfo {volume} {117}},\ \bibinfo {pages} {14220} (\bibinfo {year}
  {2013})}\BibitemShut {NoStop}%
\bibitem [{\citenamefont {Aryanpour}\ \emph
  {et~al.}(2014{\natexlab{b}})\citenamefont {Aryanpour}, \citenamefont
  {Roberts}, \citenamefont {Sandhu}, \citenamefont {Rathore}, \citenamefont
  {Shukla},\ and\ \citenamefont {Mazumdar}}]{doi:10.1021/jp410793rAryanpour}%
  \BibitemOpen
  \bibfield  {author} {\bibinfo {author} {\bibfnamefont {K.}~\bibnamefont
  {Aryanpour}}, \bibinfo {author} {\bibfnamefont {A.}~\bibnamefont {Roberts}},
  \bibinfo {author} {\bibfnamefont {A.}~\bibnamefont {Sandhu}}, \bibinfo
  {author} {\bibfnamefont {R.}~\bibnamefont {Rathore}}, \bibinfo {author}
  {\bibfnamefont {A.}~\bibnamefont {Shukla}}, \ and\ \bibinfo {author}
  {\bibfnamefont {S.}~\bibnamefont {Mazumdar}},\ }\href {\doibase
  10.1021/jp410793r} {\bibfield  {journal} {\bibinfo  {journal} {The Journal of
  Physical Chemistry C}\ }\textbf {\bibinfo {volume} {118}},\ \bibinfo {pages}
  {3331} (\bibinfo {year} {2014}{\natexlab{b}})}\BibitemShut {NoStop}%
\bibitem [{\citenamefont {Chakraborty}\ and\ \citenamefont
  {Shukla}(2014)}]{himanshu-triplet}%
  \BibitemOpen
  \bibfield  {author} {\bibinfo {author} {\bibfnamefont {H.}~\bibnamefont
  {Chakraborty}}\ and\ \bibinfo {author} {\bibfnamefont {A.}~\bibnamefont
  {Shukla}},\ }\href {\doibase http://dx.doi.org/10.1063/1.4897955} {\bibfield
  {journal} {\bibinfo  {journal} {The Journal of Chemical Physics}\ }\textbf
  {\bibinfo {volume} {141}},\ \bibinfo {eid} {164301} (\bibinfo {year}
  {2014})}\BibitemShut {NoStop}%
\bibitem [{\citenamefont {Shukla}(2002)}]{PhysRevB.65.125204Shukla65}%
  \BibitemOpen
  \bibfield  {author} {\bibinfo {author} {\bibfnamefont {A.}~\bibnamefont
  {Shukla}},\ }\href {\doibase 10.1103/PhysRevB.65.125204} {\bibfield
  {journal} {\bibinfo  {journal} {Phys. Rev. B}\ }\textbf {\bibinfo {volume}
  {65}},\ \bibinfo {pages} {125204} (\bibinfo {year} {2002})}\BibitemShut
  {NoStop}%
\bibitem [{\citenamefont {Shukla}(2004)}]{PhysRevB.69.165218Shukla69}%
  \BibitemOpen
  \bibfield  {author} {\bibinfo {author} {\bibfnamefont {A.}~\bibnamefont
  {Shukla}},\ }\href {\doibase 10.1103/PhysRevB.69.165218} {\bibfield
  {journal} {\bibinfo  {journal} {Phys. Rev. B}\ }\textbf {\bibinfo {volume}
  {69}},\ \bibinfo {pages} {165218} (\bibinfo {year} {2004})}\BibitemShut
  {NoStop}%
\bibitem [{\citenamefont {Gundra}\ and\ \citenamefont
  {Shukla}(2011{\natexlab{b}})}]{PhysRevB.84.075442Gundra}%
  \BibitemOpen
  \bibfield  {author} {\bibinfo {author} {\bibfnamefont {K.}~\bibnamefont
  {Gundra}}\ and\ \bibinfo {author} {\bibfnamefont {A.}~\bibnamefont
  {Shukla}},\ }\href {\doibase 10.1103/PhysRevB.84.075442} {\bibfield
  {journal} {\bibinfo  {journal} {Phys. Rev. B}\ }\textbf {\bibinfo {volume}
  {84}},\ \bibinfo {pages} {075442} (\bibinfo {year}
  {2011}{\natexlab{b}})}\BibitemShut {NoStop}%
\bibitem [{\citenamefont {Sony}\ and\ \citenamefont
  {Shukla}(2007)}]{sony-acene-lo}%
  \BibitemOpen
  \bibfield  {author} {\bibinfo {author} {\bibfnamefont {P.}~\bibnamefont
  {Sony}}\ and\ \bibinfo {author} {\bibfnamefont {A.}~\bibnamefont {Shukla}},\
  }\href {\doibase 10.1103/PhysRevB.75.155208} {\bibfield  {journal} {\bibinfo
  {journal} {Phys. Rev. B}\ }\textbf {\bibinfo {volume} {75}},\ \bibinfo
  {pages} {155208} (\bibinfo {year} {2007})}\BibitemShut {NoStop}%
\bibitem [{\citenamefont {Sony}\ and\ \citenamefont
  {Shukla}(2010)}]{Sony2010821}%
  \BibitemOpen
  \bibfield  {author} {\bibinfo {author} {\bibfnamefont {P.}~\bibnamefont
  {Sony}}\ and\ \bibinfo {author} {\bibfnamefont {A.}~\bibnamefont {Shukla}},\
  }\href {\doibase http://dx.doi.org/10.1016/j.cpc.2009.12.015} {\bibfield
  {journal} {\bibinfo  {journal} {Computer Physics Communications}\ }\textbf
  {\bibinfo {volume} {181}},\ \bibinfo {pages} {821 } (\bibinfo {year}
  {2010})}\BibitemShut {NoStop}%
\bibitem [{\citenamefont {Lieb}(1989)}]{Liebs-theorem}%
  \BibitemOpen
  \bibfield  {author} {\bibinfo {author} {\bibfnamefont {E.~H.}\ \bibnamefont
  {Lieb}},\ }\href {\doibase 10.1103/PhysRevLett.62.1201} {\bibfield  {journal}
  {\bibinfo  {journal} {Phys. Rev. Lett.}\ }\textbf {\bibinfo {volume} {62}},\
  \bibinfo {pages} {1201} (\bibinfo {year} {1989})}\BibitemShut {NoStop}%
\end{thebibliography}%

\end{document}